\documentclass[]{IEEEtran}

\pdfoutput=1

\newif\ifanon
\anonfalse

\newif\ifdraft
\drafttrue

\usepackage{mathtools}
\usepackage{stmaryrd}
\usepackage{float}
\usepackage{wrapfig}
\usepackage{framed}
\usepackage{xcolor}
\usepackage{xspace}
\usepackage{amssymb}
\usepackage{amsthm}
\usepackage[T1]{fontenc}
\usepackage[utf8]{inputenc}
\usepackage{bussproofs}
\usepackage[pass]{geometry}
\usepackage{url}
\usepackage{tikz}
\usepackage{framed}

\input{macros.sty}

\newtheorem{theorem}{Theorem}
\newtheorem{proposition}{Proposition}

\theoremstyle{definition}

\newtheorem{definition}{Definition}

\theoremstyle{remark}
\newtheorem{remark}{Remark}
\newtheorem{example}{Example}

\title{Formal Computational Unlinkability Proofs of RFID Protocols}

\ifanon

\author{\IEEEauthorblockN{Anonymous Submission}
\IEEEauthorblockA{ }
\IEEEauthorblockA{ }}

\else 

\author{\IEEEauthorblockN{
	Hubert Comon\IEEEauthorrefmark{1},
	Adrien Koutsos\IEEEauthorrefmark{1},
}

\IEEEauthorblockA{\IEEEauthorrefmark{1}
LSV, CNRS \& ENS Paris-Saclay},
}

\fi

\begin{document}
\maketitle

\begin{abstract}
We set up a framework for the formal proofs of RFID protocols in the computational model.
We rely on  the so-called \emph{computationally
complete symbolic attacker model}. Our contributions are:
\begin{enumerate}
\item To design (and prove sound) axioms reflecting the properties of hash functions (Collision-Resistance, PRF).
\item To formalize computational unlinkability in the model.
\item To illustrate the method, providing the first formal proofs of unlinkability of RFID protocols, in the computational model.
\end{enumerate}
\end{abstract}

\section{Introduction}

It is important  to increase our confidence in the security of protocols.
Using formal methods to prove a formal security property is the
best way to get a strong confidence. 
There is however a difficulty: we need not only to specify formally the
programs and the security properties, but also the attacker. 

One of the most popular attacker model, sometimes called the ``Dolev-Yao''
attacker, consists in assuming that, in between any message emission and the
corresponding reception, the attacker may apply a fixed set of rules modifying the
message. In addition, the attacker schedules the communications. 
More precisely, the messages are terms in a formal algebra and the rules are
given by a fixed set of combination abilities, together with a rewrite system 
specifying how to simplify the terms.  

There are several more or less automatic verification tools that rely on such a model.
Let us cite \textsf{ProVerif} \cite{proverif},  \textsf{Tamarin} \cite{Meier:2013:TPS:2526861.2526920} and \textsf{APTE} 
\cite{Cheval2014} for instance. 
Completing a proof with one of these tools will however only prove the security in the
corresponding DY model. 

Another popular attacker model, the \emph{computational attacker}, gives the same
network control to the attacker as in the Dolev-Yao model, but does not limit the attacker computations to
the combination of a fixed set of operations: any probabilistic polynomial time
computation is possible. More precisely, messages are bitstrings, random
numbers are typically bitstrings in $\{0,1\}^\eta$ (where $\eta$ is the \emph{security parameter})
and the attacker's computation time is bounded by a polynomial in $\eta$.
This model reflects more accurately a real-world attacker than the Dolev-Yao model, but formal proofs are harder to complete and more error-prone. 

There exist several formal verification tools in the computational model. For example \textsc{EasyCrypt}~\cite{Barthe2011} can be used for the construction of provably secure cryptographic primitives, and \textsc{CryptoVerif}~\cite{cryptoverif} and \textsf{F$\star$}~\cite{Barthe:2014:PRV:2535838.2535847} have been used for the study of security protocols (e.g. \cite{BlanchetJaggardScedrovTsayAsiaCCS08, 6547126}). As expected, such tools are less automatic than the verification tools in the DY model. Using such tools, we may also fail to find a proof while there is one.

We advocate the use of another approach, sketched in \cite{bana12post,bana14ccs}, which
allows to complete formal proofs in the computational model that can be automated and formally checked. 
This method has many other advantages, some of which are given in these papers: it could be applied, in principle,
to more powerful attacker models (such as attackers having access to some side-channels); it can be used to derive ``minimal'' properties
of the primitives that are sufficient to entail the security of the protocol (we will come
back to this feature later). 

The main technique, which we will recall, is to express the security of a protocol as
the unsatisfiability of a formula in first-order logic. The formula contains \emph{axioms},
reflecting the assumed computational properties of the security primitives, and the negation of the
security property, applied to terms reflecting the execution(s) of the protocol. 
This approach is, we believe, simpler than formally specifying computational security games,  probabilistic machines and simulations:
there is no security parameter, no probabilities, no timing
constraints, no Turing machines \dots Nonetheless, in case of success, the proof is valid in any model
of the axioms, including the computational model.

Compared to \textsc{EasyCrypt}, our logic works at a more abstract level, using only first-order formulas and targeting full automation and a form of completeness (saturating the axioms and the negation of the security property implies that there is an attack).

As it is now, this approach does not provide
any quantitative information: the protocol is computationally secure or it is not. 
We could however extract from a  security proof a bound on the success probability of the attacker, depending on its
computational power: we would only have to compose the adversary's advantages corresponding to each clause. 
Also, because the logic does not include the security parameter, we  can only prove the security of a number of sessions of the protocol
that does not depend on the security parameter (while there is no such limitation in 
\textsc{CryptoVerif} for instance). Still, it subsumes the symbolic approach, in which the number of sessions is also
independent of the security parameter.

A related logic for reachability security properties has been introduced in~\cite{bana12post} and is implemented in the prototype tool \textsc{Scary}~\cite{scerri15}.
It has been used for a few experiments. 
The logic for indistinguishability properties, introduced in~\cite{bana14ccs}, is not (yet) implemented. There is only one toy example
provided in~\cite{bana14ccs}, another one in \cite{DBLP:journals/iacr/BanaC16} and a more significant case study
developed in~\cite{DBLP:conf/csfw/ScerriS16}.

We investigate in this paper the application of this approach to security proofs of RFID protocols, typically proofs of
unlinkability. There are a lot of such protocols that appeared in the literature, most of which are very simple, due to
the low computing capabilities of a RFID tag: the protocols often use only primitives such as hashing, xoring, pairing and splitting. These protocols have been studied, attacked, patched and automatically proved in the 
DY model (see for instance \cite{hirschi16sp}).  On the computational side, \cite{Vaudenay2007} investigates the 
computational definitions of unlinkability,
together with examples of RFID protocols that satisfy (or not) the definitions.  There are however very few proofs of security
in the computational model and no (up to our knowledge) \emph{formal} security proof. For instance, an RFID protocol is proposed
in \cite{DBLP:conf/ccs/LeBM07}, together with a universally composable (claimed) proof. The proof is however quite informal, and 
attacks were found on this protocol (see~\cite{DBLP:conf/ispec/OuafiP08}). Admittedly, such attacks can be easily circumvented, but this shows
that a formal approach is useful, if not necessary. Similarly, as reported in \cite{Juels:2009:DSP:1609956.1609963},
other RFID  protocols that were claimed secure turned out to be broken.

A large fraction of RFID protocols, the so-called \emph{Ultralightweight} RFID protocols (e.g. \cite{Chien:2007:SNU:1320305.1320890, DBLP:conf/wisa/Peris-LopezCER08}), aim at ensuring only weak security properties, and against passive attackers, because of the strong constraints on the number of gates embedded in the RFID tags. We do not consider such protocols in this paper.


\paragraph{Contributions} The contributions of this paper are:
\begin{enumerate}
\item To design axioms that reflect security assumptions on the primitives that are used in the RFID protocols
(typically hash functions, pseudo-random generators and xor), and to prove their correctness.
\item To express formally the computational unlinkability. There are various definitions; we chose to formalize one of them (from \cite{Juels:2009:DSP:1609956.1609963}).
Most other definitions can be expressed in a similar way. The security property is expressed as the indistinguishability of two sequences of terms. These terms are computed from the protocol specification extended with corruption capabilities. We use a specific technique inspired by the folding of transition systems described in~\cite{bana14ccs}.
\item To illustrate the proof technique on two examples taken from \cite{van2008attacks}: \kcl and \lak.
As far as we know, all published RFID protocols, that do not rely on encryption, are computationally insecure. 
This is also the case of these two protocols.
We propose small modifications of the protocols, which prevent the known
attacks. Some of the modified versions are secure in the DY model. Depending on the assumptions
on the primitives, they may however be insecure in the computational model. For instance, if we assume
the  hash function to be pre-image resistant and one-way, the corrected version of \lak,
proved in \cite{hirschi16sp}, is not necessarily computationally secure: there might be attacks on both 
authentication and unlinkability. We actually need a family of keyed hash functions, which satisfies
the \emph{pseudo-random functions} (PRF)  property. With the appropriate implementation assumptions, we formally 
prove the security
of the two protocols. Up to our knowledge, these are the first formal security proofs of RFID protocols in the computational
model. 
\end{enumerate}

\paragraph{Outline} In Section~\ref{section:axioms}, we briefly recall the methodology described in
\cite{bana14ccs} and we propose some axioms for the hashing and
exclusive or, that depend on the assumptions on the cryptographic libraries. 
In Section~\ref{section:security} we recall the definition of privacy of a RFID protocol given by Juels and Weis in~\cite{Juels:2009:DSP:1609956.1609963}, and we show how this property translates in the logic. 
In Section~\ref{section:protocols} we recall the two protocols \kcl and \lak, known attacks on them and formally prove the
security of fixed versions of the protocols. We also show that relaxing the assumptions yields some attacks.
Finally, in Section \ref{section:pseudo-random}, we show (as expected) that abstracting pseudo-random numbers with
random numbers is sound, provided that the seed is not used for any other purpose.




\section{The logic}
\label{section:axioms}

Our goal is to formally study the protocols in the computational model.
In order to do this we follow the directions described in \cite{bana14ccs}: we specify in a first-order
logic what the attacker \emph{cannot do}, which yields a set of axioms $A$. We 
also compute from the protocol and the security property a formula $\neg\psi$ expressing that
there is an attack on the protocol. We know that if $A\cup\{ \neg \psi\}$ is unsatisfiable, then the
protocol is secure in any model of $A$. If, in addition, every axiom is computationally sound,
then the protocol is computationally secure. 

In this section we recall the first-order (indistinguishability) logic and provide a set
of axioms $A$, some of which are valid in any computational model while others require
some security assumptions on the cryptographic primitives.

\subsection{Syntax of the logic}

\paragraph{Terms} In the logic terms are built on a set of function symbols ${\cal F}$, a set of function symbols $\cal G$  (used to represent the attacker's computations), a set of names $\cal N$ and a set of variables $\cal X$. 

In the examples that are considered in this paper, ${\cal F}$ contains at least the following function symbols:
\[\hash,\xor, \pair{\_}{\_}, \eq{\_}{\_}, \ite{\_}{\_}{\_},\true,\false, \pi_1,\pi_2,\textbf{0}\]

Each variable and term has a \emph{sort}, which is either \bool, \Nonce, or \term.  Every term of sort \bool~or \Nonce, has also the sort \term. The typing rules for function symbols are defined as follows:
\begin{itemize}
\item $\pair{\_}{\_}: \term\times \term \rightarrow \term$
\item $\true,\false: \rightarrow \bool$
\item $\pi_1,\pi_2:\term \rightarrow \term$
\item $\eq{\_}{\_} : \term \times \term \rightarrow \bool$

\item Names have type $\Nonce$

\item In the computational interpretation of the terms, we will only need to xor bitstrings of a fixed length: the length of
the random numbers. We therefore restrict the type of $\xor$ to: \(\Nonce \;\times \;\Nonce \;\rightarrow\; \Nonce\)

\item $\ite{\_}{\_}{\_}$ has three types:\\
$\begin{array}{l}
\bool\times \term \times \term  \rightarrow \term\\
\bool\times\bool\times \bool  \rightarrow  \bool\\
\bool \times \Nonce \times \Nonce \rightarrow  \Nonce
\end{array}$

\item In our examples, hash values will have a length equal to the security parameter, so 
$\hash$ returns a value of sort $\Nonce$. Furthermore, we need a \emph{keyed} hash function, otherwise
we cannot state any computationally sound property over the hash function $\hash$: \(\term \times \Nonce \rightarrow \Nonce \)
\item $\textbf{0}:\rightarrow \Nonce$
\end{itemize}

We can also use arbitrary additional symbols in $\cal F$ if needed. 
 Each term has always a least sort (w.r.t. the ordering $\bool,\Nonce < \term$), which we call the \emph{sort of the term}.\\

\paragraph{Formulas} Atomic formulas aim at representing the indistinguishability of two experiment. They are expressions :
\[ u_1,\ldots,u_n \sim v_1,\ldots,v_n\]
where $u_1,\ldots,u_n,v_1,\ldots,v_n$ are terms and, for every $i$, $u_i$ and $v_i$ have the same sort. More formally, for every $n$, we are using a predicate symbol $\sim_n$ with $2n$ arguments. Formulas then are obtained by combining atomic formulas with Boolean connectives $\wedge, \rightarrow, \vee, \neg$
and (first-order) quantifiers.

\paragraph{Syntactic shorthands}
Displaying the formulas and axioms, we use the notation $\vec{u}$ for  a sequence of terms of the appropriate type.
We also use ``$=$'' as a syntactic sugar: $u=v$ is a shorthand for $\eq{u}{v}\sim \true$.
We may also use logical connectives in the conditions, as a shorthand. For instance 
``$\ite{b_1\vee b_2}{x}{y}$''  is a shorthand for ``$\ite{b_1}{x}{\ite{b_2}{x}{y}}$''.

\subsection{Semantics of the logic}

We rely on classical first-order interpretations: function symbols are interpreted as functions on an underlying
interpretation domain and predicate symbols are interpreted as relations on this domain.

Though, we assume that the interpretation satisfies the generic axioms of the Figure~\ref{figure:generic-axioms}, which
do not depend on the actual cryptographic libraries. These axioms are computationally valid, as we will see.

One particular class of interpretations of the logic are the so-called \emph{computational semantics}.
We recall here the main features (the complete definition can be found in \cite{bana14ccs}): 
\begin{itemize}
\item The domain of a computational model  is 
the set of deterministic polynomial time Turing machines equipped with an input (and working) tape and two random tapes
(one for honestly generated random values, the other for random values directly available to the attacker). The length of
the machine input is the security parameter $\eta$.
\item A name $n\in {\cal N}$ is interpreted as a machine that, on input of length $\eta$, extracts a word of length $\eta$
from the first random tape $\rho_1$. Furthermore, different names extract disjoint parts of $\rho_1$.
\item $\true$, 
$\false$,  $\eq{\_}{\_}$,  
$\ite{\_}{\_}{\_}$, $\xor$
are always interpreted in the expected way. For instance,
$\ite{\_}{\_}{\_}$, takes three machines $M_1,M_2,M_3$ and returns a machine $M$ such that, on input $w,\rho_1.\rho_2$,  $M$ returns $M_2(w,\rho_1,\rho_2)$ if
$M_1(w,\rho_1,\rho_2)=1$ and returns $M_3(w,\rho_1,\rho_2)$ otherwise. 
\item Other function symbols $f$ in $\cal F$ (including $\hash$ and $\pair{\_}{\_}$, $\pi_1,\pi_2$) are interpreted as arbitrary deterministic polynomial
time Turing machines. Though, we may restrict the class of interpretations using assumptions on the implementations of
the primitives. This is discussed in the next section.
\item A function symbol $g\in {\cal G}$ with $n$ arguments is interpreted as a function $\sem{g}$, such that there is a polynomial time Turing machine ${\cal A}_g$ such that for every $n$ machines $d_1,\ldots,d_n$ in the interpretation domains, and for every inputs $w,\rho_1,\rho_2$,
\begin{multline*}
 \sem{g}(d_1,\ldots,d_n)(w,\rho_1,\rho_2) =
\\ {\cal A}_g(d_1(w,\rho_1,\rho_2),\ldots,d_n(w,\rho_1,\rho_2),\rho_2)
\end{multline*}
In particular, the machine does not use directly the random tape $\rho_1$.
\item The interpretation is lifted to terms: given an assignment $\sigma$ of the variables of a term $t$ to the domain values, we write $\sem{t}_{\eta,\rho_1,\rho_2}^\sigma$ the computational interpretation of the term, with respect to $\eta,\rho_1,\rho_2$.
The assignment $\sigma$ is omitted when empty. We may also omit the other parameters when they are irrelevant.
\item The predicates $\sim$ are interpreted as \emph{computational indistinguishability} $\approx$, defined by:
\[ d_1,\ldots,d_n \approx d'_1,\ldots,d'_n\]
iff, for every deterministic polynomial time Turing machine $\mathcal{A}$,
\begin{align*}
\big|&\prob(\rho_1,\rho_2: \; {\cal A}(d_1(1^\eta,\rho_1,\rho_2),\ldots,d_n(1^\eta,\rho_1,\rho_2))=1) -\\
&\prob(\rho_1,\rho_2: \; {\cal A}(d'_1(1^\eta,\rho_1,\rho_2),\ldots,d'_n(1^\eta,\rho_1,\rho_2))=1) \big|
\end{align*}
is negligible when $\rho_1,\rho_2$ are drawn according to the uniform distribution. 
\end{itemize}

\subsection{Computationally valid axioms}

We only consider the purely universal fragment of first-order logic: every variable is implicitly universally quantified. A formula is \emph{computationally valid} if it is valid in all computational interpretations.

\begin{figure}
\noindent\framebox{\parbox{0.98\linewidth}{
\begin{itemize}
\item $(Refl), (Sym), (Trans):$ $\sim$ is reflexive, symmetric and transitive. 

\item For all $\vec u, \vec v,t,t'$ of appropriate types:
\[\tag{$Dup$}  \vec u,t \sim \vec v,t \quad\Rightarrow \quad \vec u,t,t \sim \vec v,t',t'\]

\item $(Congr):$  $=$ is a congruence

\item For any permutation $\pi$ of $1,\ldots, n$ and for all $x_1,\ldots,x_n,y_1,\ldots,y_n$ of appropriate types:
\begin{multline*} 
\tag{$Perm$}
x_1\ldots,x_n \sim y_1,\ldots, y_n \\\Rightarrow x_{\pi(1)},\ldots,x_{\pi(n)} \sim y_{\pi(1)},\ldots,y_{\pi(n)}
\end{multline*}

\item For every function symbol $f$ of appropriate type, for all $\vec{x},\vec{y},\vec{x'},\vec{y'}$:
\[ \tag{$FA$}  \vec{x},\vec{y}\sim \vec{x'}, \vec{y'} \quad \Rightarrow \quad f(\vec{x}),\vec{y} \sim f(\vec{x'}), \vec{y'} \]

\item For every $b,x,y,z,\vec{w}$ of appropriate types:
\end{itemize}
  \[ 
    \begin{array}{c}
      \begin{array}{rcl}
        \eq{x}{x} & = & \true \\
        \ite{\true}{x}{y} & = & x\\
        \ite{\false}{x}{y} & = & y\\
        \ite{b}{x}{x}& = & x\\
        \ite{b}{\ite{b}{x}{y}}{z} & = & \ite{b}{x}{z}\\
        \ite{b}{x}{\ite{b}{y}{z}}& = & \ite{b}{x}{z}
      \end{array}\\
      \begin{array}{l}
        \ite{b}{(\ite{a}{x}{y})}{z} =\\
        \phantom{aaa}\null\hfill \ite{a}{(\ite{b}{x}{z})}{(\ite{b}{y}{z})}\\
        \ite{b}{x}{(\ite{a}{y}{z})} =\\
        \null\hfill \ite{a}{(\ite{b}{x}{y})}{(\ite{b}{x}{z})}\\
        f(\vec{x},\ite{b}{y}{z},\vec{w}) =\\
        \null\hfill\ite{b}{f(\vec{x},y,\vec{w})}{f(\vec{x},z,\vec{w})}
      \end{array}\\
      \begin{array}{rcl}
        x\xor y & = & y\xor x\\
        x\xor x & = & \textbf{0}\\
        \textbf{0}\xor x & = & x\\
        x \xor (y\xor z) & = & (x\xor y)\xor z \\
      \end{array}
    \end{array}
\]
\begin{itemize}
\item If $\nonce$ does not occur in $x,y,\vec{u},\vec{v}$:
\begin{gather*}
  \tag{$Indep$} \vec u \sim \vec v \Rightarrow   \vec{u}, x \xor \nonce \sim \vec{v}, y \xor \nonce \\
 \tag{$EqIndep$} \eq{\nonce}{x} =\false
\end{gather*}

\item If $\nonce$ does not occur in $\vec{u}$ and $\nonce'$ does not occur in $\vec{u'}$:
\[ \tag{$FreshNonce$}  \vec{u}\sim \vec{u'}\Rightarrow \vec{u},\nonce \sim \vec{u'},\nonce' \]

\item For every $x,y,z,\vec{w},\vec{x'}$ of appropriate types and for every context $t$:
 \begin{multline*}
 \tag{$IfThen$} \ite{\eq{x}{y}}{t[x]}{z},\vec{w} \sim \vec{x'}  \\
\Rightarrow \ite{\eq{x}{y}}{t[y]}{z}, \vec{w}\sim \vec{x'}
\end{multline*}

\item $(CS):$ For every $b,b',x,x',y,y',z,z',\vec u, \vec u'$ of appropriate types:
\begin{gather*}
 b,\ite{b}{x}{z},\vec{u}\sim b',\ite{b'}{x'}{z'},\vec u' \\
\wedge  b,\ite{b}{z}{y},\vec{u}\sim b',\ite{b'}{z'}{y'},\vec u'\\
\Rightarrow\ite{b}{x}{y},\vec{u}\sim \ite{b'}{x'}{y'}, \vec u'
\end{gather*}
\end{itemize}
}}
\caption{Axioms that are independent of the cryptographic primitives}
\label{figure:generic-axioms}
\end{figure}

\begin{proposition}\label{proposition:generic:axioms}
The axioms displayed in the figure \ref{figure:generic-axioms} are computationally valid.
\end{proposition}

Most of the axioms have already been proven valid elsewhere (for instance in \cite{bana14ccs,DBLP:journals/iacr/BanaC16}).
Only the axioms $\vec{u}, x \xor \nonce \sim \vec{u}, y \xor \nonce$ and $\eq{\nonce}{x} =\false$ are new, but their computational validity is quite straightforward to prove (e.g. for every $x$, the distributions $x\xor \nonce$ and $y \xor \nonce$ are both uniform distributions). \textsc{EasyCrypt} is using a similar rule through the \textsf{rnd} tactic.
\ifdraft
A formal proof of these two axioms is given in Appendix~\ref{appendix:generic-axioms}.
\else \fi

\subsection{Assumptions on primitives}
Some of our axioms reflect implementation assumptions, that is, identities that must be satisfied by any implementation of these functions. For example we will assume  that the projections of a pair return the components of the pair:
\[ \pi_1(\pair{x}{y})= x \qquad \pi_2(\pair{x}{y})=y \]
We do not make any further assumption on the implementation of pairing.

Other axioms reflect cryptographic assumptions  on the primitives. For example, for hash functions, we need to express some computational hardness properties. An example of such a property is the Collision Resistance property, that we recall below:
 \begin{definition}[CR-HK \cite{Goldwasser01lecturenotes}] 
   A hash function $\hash$ is said to be \emph{collision resistant under hidden-key attacks}
   if and only if for all probabilistic polynomial time adversary $\mathcal{A}$ with access to an oracle returning the keyed hash value of the request,
   \[ \prob(\key,\rho: \; \mathcal{A}^{\hash(\cdot,\key)}(1^\eta) = \pair{m_1}{m_2} \wedge \hash(m_1,\key) = \hash(m_2,\key) )\]
   is negligible, where $\key$ is drawn uniformly at random in $\{0,1\}^\eta$ and $\rho$ is the random tape of the attacker.
 \end{definition}

This property can be expressed in the logic by the following $CR$ axiom:\\

\noindent\framebox{\parbox{0.98\linewidth}{
If the only occurrences of $\key$  in $t, t',\vec u$ are as a second argument of $\hash$:
\begin{gather*}
 \vec u,\ite{\eq{t}{t'}}{\false}{\eq{\hash(t,\key)}{\hash(t',\key)}} \\
\sim\\ 
\vec u, \false
\end{gather*}
}}

\begin{remark} 
Note that we need the conditional here: we cannot simply state that when $t$ and $t'$ are distinct,
 \[ \eq{\hash(t,\key)}{\hash(t',\key)}\sim\false\]

For instance, take $t= g(u)$ and $t'=g(u')$ where $u,u'$ are distinct and $g$ is an attacker's function. Then even though $t$ and $t'$ are syntactically
distinct, the function $g$ can be interpreted as a function that discards its argument and always return the same value.
In such a case, the computational interpretations of $t$ and $t'$ are identical.
\end{remark}

\begin{proposition}\label{prop:crhk}
The $CR$  axiom is sound in any computational model \cmodel where the hash function $\hash$ is collision resistant under hidden-key attacks.
\end{proposition}

\begin{IEEEproof}
Let $b_{CR}$ be the following term:
\[
  \ite{\eq{t}{t'}}{\false}{\eq{\hash(t,\key)}{\hash(t',\key)}}
\]
Let \cmodel be a computational model such that $\hash$ is interpreted by a  collision-resistant keyed hash function, and assume that there exists an adversary $\mathcal{A}$ such that:
\begin{equation*}
\big|\Pr\left(\rho : \mathcal{A}(\llbracket b_{CR}\rrbracket_{\eta,\rho})\right) 
-\Pr\left(\rho : \mathcal{A}(\llbracket\false\rrbracket_{\eta,\rho})\right) \big|
\end{equation*}
is not negligible in $\eta$. Since $b_{CR}$ is a boolean term, $\sem{b_{CR}}_{\eta,\rho}\in \{\true,\false\}$, hence the existence of $\mathcal{A}$ is equivalent to
\begin{quote} $\Pr\left(\rho : \llbracket b_{CR}\rrbracket_{\eta,\rho} = \true\right)$  is non-negligible
\end{quote}
We are going to define a probabilistic polynomial time adversary with access to an hash oracle with a non negligible chance of winning the CR-HK game. 

Since the only occurrences of $\key$ in $t, t',\vec u$ are as a second argument of $\hash$, we can define $\mathcal{A}'$ to be the CR-HK-adversary that computes and returns $\llbracket t \rrbracket_{\eta,\rho}$ and $\llbracket t' \rrbracket_{\eta,\rho}$ (names different from $\key$ are computed by uniform sampling, and subterms of the form $\hash( u,\key)$ are computed by calling the hash oracle on the inductively computed $\llbracket u \rrbracket_{\eta,\rho}$). Then we have:
\[
\begin{array}{llr}
&\lefteqn{ \Pr\big(\key :\; \mathcal{A}'^{\hash(\cdot,\key)}() = \pair{m_1}{m_2}}\qquad\\
&& \wedge \hash(m_1,\key) = \hash(m_2,\key) \wedge m_1 \ne m_2\big)\\
=& \lefteqn{\Pr\big(\rho:\; \pair{\llbracket t \rrbracket_{\eta,\rho}}{\llbracket t' \rrbracket_{\eta,\rho}} = \pair{m_1}{m_2}  \wedge}\\
&& \hash(m_1,\llbracket\key\rrbracket_{\eta,\rho}) = \hash(m_2,\llbracket\key\rrbracket_{\eta,\rho}) \wedge m_1 \ne m_2 \big)\\
=&\lefteqn{ \Pr\big(\rho:\;\hash(\llbracket t \rrbracket_{\eta,\rho},\llbracket\key\rrbracket_{\eta,\rho}) = \hash(\llbracket t \rrbracket_{\eta,\rho},\llbracket\key\rrbracket_{\eta,\rho})}\\
&&\wedge \llbracket t \rrbracket_{\eta,\rho} \ne \llbracket t' \rrbracket_{\eta,\rho} \big)\\
=& \lefteqn{\Pr\big(\rho:\,\llbracket \hash(t,\key) \rrbracket_{\eta,\rho} = \llbracket \hash(t',\key) \rrbracket_{\eta,\rho} \wedge \llbracket t \rrbracket_{\eta,\rho} \ne \llbracket t' \rrbracket_{\eta,\rho}\big)}\\
=&\lefteqn{\Pr \big(\rho:\; \llbracket b_{CR}\rrbracket_{\eta,\rho} = \true\big)}
\end{array}
\]
which we assumed to be non negligible in $\eta$.


\end{IEEEproof}

Unfortunately, this property is not sufficient to prove the security of the RFID protocols considered in this paper (we will show attacks when only CR-HK is assumed). Therefore we need to consider a stronger property:
\begin{definition}[PRF\cite{DBLP:books/cu/Goldreich2001,DBLP:journals/jacm/GoldreichGM86}] 
  A family of keyed hash functions: $H(\cdot,k): \{0,1\}^* \rightarrow \{0,1\}^\eta$ is a \emph{Pseudo Random Function} family
  if, for any PPT adversary $\cal A$ with access to an oracle ${\cal O}_f$:
  \[ |\prob(k,\rho: \; {\cal A}^{{\cal O}_{H(\cdot,k)}}(1^\eta)=1) - \prob(g,\rho:\; {\cal A}^{{\cal O}_{g(\cdot)}}(1^\eta)=1)|\]
  is negligible.
  Where  
  \begin{itemize}
  \item $k$ is drawn uniformly in $\{0,1\}^\eta$.
  \item $\rho$ is the random tape of the attacker $\cal A$ (drawn uniformly).
  \item $g$ is drawn uniformly in the set of all functions from $\{0,1\}^*$ to $\{0,1\}^\eta$.
  \end{itemize}
\end{definition}

We express this property in the logic using the following recursive schema of axioms:\\

\noindent\framebox{\parbox{0.98\linewidth}{
\[  \vec{u}, \ite{c}{\textbf{0}}{\hash(t,k)} \sim \vec{u}, \ite{c}{\textbf{0}}{\nonce} \quad (PRF_n)
\]
where:
\begin{itemize}
\item the occurrences of $\hash$ (and $\key$) in $\vec{u},{t}$ are $\hash(t_1,\key),\ldots,\hash(t_n,\key)$
\item $\nonce$ is a name, that does not occur in $\vec{u},t,c$
\item $c\equiv \bigvee_{i=1}^n \eq{t_i}{t}$.
\end{itemize}
}}

\begin{proposition}\label{proposition:prfn}
For any $n$, $(PRF_n)$ is computationally sound  if $\hash(\cdot,k)$ is a PRF family.
\end{proposition}

\ifdraft
The proof can be found in Appendix~\ref{appendix:prf},
\else
The proof can be found in~\cite{full-version},
\fi
and is a little more complicated than the previous one. 
Note that $\key$ cannot appear in $t_i$ itself, unless  $\hash(t_j,\key)$ is a subterm of $t_i$ for some $j$;
the statement of the axiom does not cover ``key cycles'' such as in the expression $\hash(\key,\key)$. Extending
the axiom to cover such situations would require a Random Oracle assumption for the hash function.


\section{Security properties}
\label{section:security}

Radio Frequency IDentification (\rfid) systems allow to wirelessly identify objects. These systems are composed of \emph{readers} and \emph{tags}. Readers are radio-transmiters connected through a secure channel to a single server hosting a database with all the tracked objects information. Tags are wireless transponders attached to physical objects
that have a limited memory and computational capacity (so as to reduce costs). For the sake of simplicity we will assume that there is only one reader, which will represent the database server as well as all the physical radio-transmitters.

\begin{example}
\label{example:kcl}
In order to illustrate the properties, we introduce  a very simple version of the  protocol \kcl. 
The original protocol from~\cite{4279773} is (informally) described in Figure~\ref{figure:kcl}. 
The key $\key_{\textsf{A}}$ is a shared secret key between the tag $T_{\textsf{A}}$ and the reader $R$. 
For simplicity we assume, for now, that names $\nonce_T,\nonce_R$ are randomly generated by each party ($\rsample$ is used to denote random sampling); this will be
justified in Section \ref{section:prng}.
The protocol is expected to ensure both authentication and unlinkability.
\begin{figure}
\[
  \begin{array}{l}
    \begin{array}{lcl}
      R&:& \nonce_{R} \rsample\\
      T_{\textsf{A}}&:& \nonce_{T} \rsample\\
    \end{array} \\[2em]
    \begin{array}{lcl}
      1: R \longrightarrow T_{\textsf{A}} &:& \nonce_R\\
      2: T_{\textsf{A}} \longrightarrow R &:& \pair{\textsf{A} \oplus \nonce_T}{\nonce_T \oplus \hash(\nonce_R,\key_\textsf{A})}
    \end{array}
  \end{array}
\]
\caption{The original \textsc{kcl} protocol}
 \label{figure:kcl}
\end{figure}
\end{example}

\subsection{Privacy of RFID protocols}

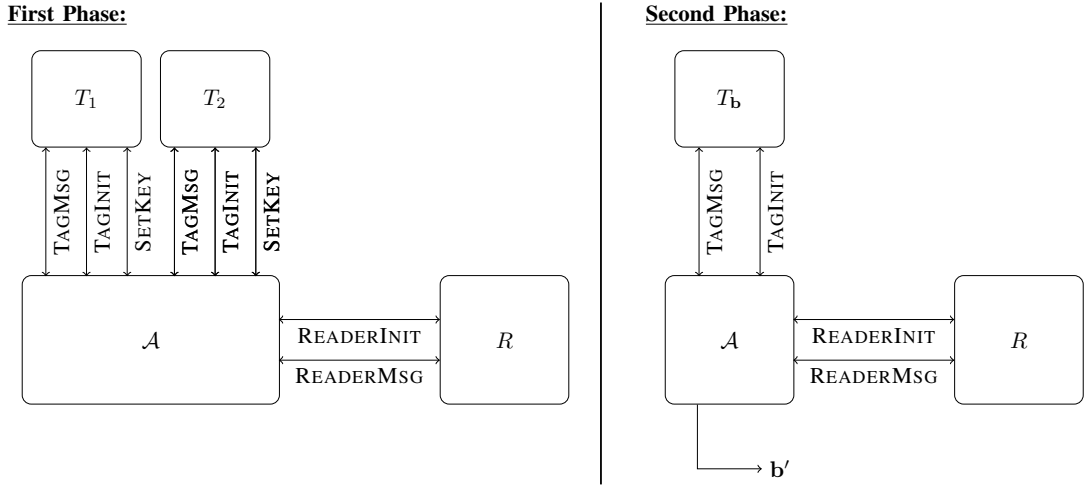
\begin{figure*}[h]
\begin{center}
\resizebox{0.8\linewidth}{!}{
\begin{tikzpicture}
\tikzset{tag/.style={draw,rounded corners,minimum height=1.5cm,minimum width=1.7cm}};
\tikzset{reader/.style={draw,rounded corners,minimum height=2cm,minimum width=2cm}};
\tikzset{adversary/.style={draw,rounded corners,minimum height=2cm,minimum width=4cm}};

\begin{scope}[shift={(-4.5,0)}]

\node at (-0.3,3.3){\underline{\textbf{First Phase:}}};

\node[tag] (a) at (0,2){$T_1$};
\node[tag] (b) at (2,2){$T_2$};
\node[adversary] (d) at (1,-1.75){$\mathcal{A}$};
\node[reader] (c) at (6.5,-1.75){$R$};

\draw[transform canvas={xshift=4ex},<->] (a.south) -- ++(0,-2) node[midway,rotate=90,below]{\textsc{SetKey}};
\draw[transform canvas={xshift=0ex},<->] (a.south) -- ++(0,-2) node[midway,rotate=90,below]{\textsc{TagInit}};
\draw[transform canvas={xshift=-4ex},<->] (a.south) -- ++(0,-2) node[midway,rotate=90,below]{\textsc{TagMsg}};

\draw[transform canvas={xshift=4ex},<->] (b.south) -- ++(0,-2) node[midway,rotate=90,below]{\textsc{SetKey}};
\draw[transform canvas={xshift=0ex},<->] (b.south) -- ++(0,-2) node[midway,rotate=90,below]{\textsc{TagInit}};
\draw[transform canvas={xshift=-4ex},<->] (b.south) -- ++(0,-2) node[midway,rotate=90,below]{\textsc{TagMsg}};

\draw[transform canvas={yshift=2ex},<->] (c.west) -- ++(-2.5,0) node[midway,below]{\textsc{ReaderInit}};
\draw[transform canvas={yshift=-2ex},<->] (c.west) -- ++(-2.5,0) node[midway,below]{\textsc{ReaderMsg}};
\end{scope}

\draw[thick] (3.5,-4) -- ++(0,7.5);

\begin{scope}[shift={(4.5,0)}]

\node at (0.8,3.3){\underline{\textbf{Second Phase:}}};

\node[tag] (a) at (1,2){$T_{\mathbf{b}}$};
\node[adversary,minimum width=2cm] (d) at (1,-1.75){$\mathcal{A}$};
\node[reader] (c) at (5.5,-1.75){$R$};

\draw[transform canvas={xshift=3ex},<->] (a.south) -- ++(0,-2) node[midway,rotate=90,below]{\textsc{TagInit}};
\draw[transform canvas={xshift=-3ex},<->] (a.south) -- ++(0,-2) node[midway,rotate=90,below]{\textsc{TagMsg}};

\draw[transform canvas={xshift=4ex},<->] (b.south) -- ++(0,-2) node[midway,rotate=90,below]{\textsc{SetKey}};
\draw[transform canvas={xshift=0ex},<->] (b.south) -- ++(0,-2) node[midway,rotate=90,below]{\textsc{TagInit}};
\draw[transform canvas={xshift=-4ex},<->] (b.south) -- ++(0,-2) node[midway,rotate=90,below]{\textsc{TagMsg}};

\draw[transform canvas={yshift=2ex},<->] (c.west) -- ++(-2.5,0) node[midway,below]{\textsc{ReaderInit}};
\draw[transform canvas={yshift=-2ex},<->] (c.west) -- ++(-2.5,0) node[midway,below]{\textsc{ReaderMsg}};

\draw[->] (d.south) ++(-0.5,0) |- ++(1,-1) node[right]{$\mathbf{b'}$};
\end{scope}
\end{tikzpicture}}
\end{center}
\caption{\label{fig:strongprivacy} Privacy game with two tags $T_1,T_2$. The adversary $\mathcal{A}$ wins if $\mathbf{b} = \mathbf{b'}$.}
\end{figure*}

We use the notion of Privacy for RFID protocols defined in~\cite{Juels:2009:DSP:1609956.1609963}, that we informally recall here. This is a game-based definition, in which the adversary is a probabilistic polynomial time Turing machine and interacts with a reader $R$ and a finite set of tags $\{T_1,\dots,T_n\}$ (also probabilistic Turing machines) through  fixed communication interfaces, which are informally described below and in Figure~\ref{fig:strongprivacy}:
\begin{itemize}
\item A tag $T_i$ can store a secret key $\key_i$, an id $Id_i$, a session identifier $sid$ and the previous $l$ challenge-response pairs of the current session. It has the following interface:
\begin{itemize}
\item \textsc{SetKey}: Corrupts the tag by returning its old key $\key_i$ and id $Id_i$, and allows the adversary to send a new key $\key'_i$ and a new id $Id_i'$ of its choice.
\item \textsc{TagInit}: Initialize a tag with a session identifier $sid'$. The tag deletes the previous session identifier and the logged challenge-response pairs.
\item \textsc{TagMsg}: The tag receives a challenge $c_i$ and returns a response $r_i$ (that was computed using the key, the session identifier and the logged challenge-response pairs). Additionally the tag logs the challenge-response pair $(c_i,r_i)$.
\end{itemize}
\item The reader $R$ stores some private key material (for example a master secret key, the tags private keys \dots) and entries of the form $(sid,status,c_0,r_0,\dots,c_l)$ where $status$ is either $open$ or $closed$ depending on whether the session is completed or on-going. It has the following interface:
\begin{itemize}
\item \textsc{ReaderInit}: Returns a fresh session identifier $sid$ along with the first challenge $c_o$. The reader also stores a new entry of the form $(sid,open,c_o)$.
\item \textsc{ReaderMsg}: The reader receives a session identifier $sid$ and a response $r_i$. It looks for a data entry of the form $(sid,open,c_0,r_0,\dots,c_i)$, appends the message $r_i$ to the data entry, and either closes the session (by changing the status from $open$ to $closed$) and outputs a new challenge message $c_{i+1}$ (possibly $0$) and appends it to the data entry.
\end{itemize}
\end{itemize}

The adversary is allowed to corrupt (by a \textsc{SetKey} command) up to $n-2$ tags. At the end of a first phase of computations and interactions with the reader $R$ and tags $\{T_1,\dots,T_n\}$, the adversary chooses two uncorrupted tags $T_{i_0}$ and $T_{i_1}$, which are removed from the set of available tags. Then one of these tags is chosen uniformly at random by sampling a bit $b$ and made accessible to the adversary as an oracle. The adversary performs a second phase of computations and interactions with the reader $R$, the tags $\{T_1,\dots,T_n\} \backslash \{T_{i_0},T_{i_1}\}$, as well as the randomly selected tag $T_{i_b}$ (obviously the adversary is not allowed to corrupt $T_{i_b}$). Finally the adversary outputs a bit $b'$, and wins if it guessed the chosen tag (that is if $b = b'$). A protocol is said to verify $m$-Privacy if any adversary $\mathcal{A}$ using at most $m$ calls to the interfaces, has a probability of winning the game bounded by $\frac{1}{2} + f_{\mathcal{A}}(\eta)$, where $f_{\mathcal{A}}$ is a negligible function in the security parameter.

\begin{remark}
Our definition of privacy is slightly different from the one in~\cite{Juels:2009:DSP:1609956.1609963}:
\begin{itemize}
\item We do not assume that the reader answers ``reject'' or ``accept'' when a session is completed. 
We can easily encode this feature by adding an answer from the reader at the end of the protocol with the corresponding message. Not taking this as the default behavior allows to model adversaries that are less powerful and do not have access to the result of the protocol.
\item We use $m$-Privacy, whereas they use $(r,s,t)$-Privacy where $r$ and $s$ are a bound on the number of calls to \textsc{ReaderInit} and \textsc{TagInit} respectively, and $t$ is a bound on the running time. We dropped the explicit mention of $t$ as we are only interested in proving privacy against any polynomial time adversary. Moreover using $m$ or $r,s$ is equivalent, as, for a given protocol, the number of calls to the interfaces is bounded by the number of calls to \textsc{ReaderInit} and to \textsc{TagInit}, and conversely.
\end{itemize}
\end{remark}

\subsection{Bounded Session Privacy}

\paragraph{Protocol Description} We let $\Gamma_n$ be the set of possible actions of an adversary interacting with $n$ tags:
\begin{multline*}
\Gamma_n = \{\textsc{SetKey}_i,\textsc{TagInit}_i,\textsc{TagMsg}_i,\\\textsc{ReaderInit},\textsc{ReaderMsg}\,|\, 1 \le i \le n \}
\end{multline*}
We are going to use a substitution $\sigma$ to save the state of the reader $R$ and the tags $T_1,\dots,T_n$  internal memories. For instance,
for each tag $T_i$, we use two locations $x_{\key_i},x_{Id_i}$ to store respectively the values of the key and of the id.

For every adversarial frame $\phi$, for every action $\alpha \in \Gamma$ and for every internal memory state $\sigma$ we can build from the protocol description a term $t^{ \sigma}_\alpha(\phi)$, which represents the answer of the reader or tag to the action $\alpha$. We also build a internal memory update $\theta^\sigma_\alpha(\phi)$ such that $\sigma\,\theta^\sigma_\alpha(\phi)$ represents the updated memory of the reader and tags after the action has been performed. We also build from the protocol description the initial internal memory $\sigma_{\text{init}}$. 

For the sake of simplicity, we  assume that all names appearing in a term $t^{ \sigma}_\alpha(\phi)$ are fresh. If one wants to reuse a name (e.g. in the \lak protocol the reader uses the same name in the first and second challenge), then we will store it in the internal memory $\sigma$.

\paragraph{Folding of the Protocol} Given a subset of actions $S$ of $\Gamma_n$,
we construct a term $t_S^\sigma(\phi)$ and a substitution $\theta_S^\sigma(\phi)$ 
representing (respectively) the message  sent over the network and the memory update, depending on the action chosen by the attacker in $S$ (under memory $\sigma$ and frame $\phi$). Intuitively, such terms gather together all possible interleavings of actions using the folding technique explained in \cite{bana14ccs}. 

Given an (arbitrary) enumeration of $S=\{\alpha_1,\ldots,\alpha_r\}$, $t_S^\sigma(\phi)$ and $\theta_S^\sigma(\phi)$ are defined using intermediate terms $u_i^\sigma(\phi)$ and $\tau_i^\sigma(\phi)$ (for all $i \in \{1,\dots,r\}$), relying on an attacker function symbol $to\in \mathcal{G}$ (representing the scheduling choice of the attacker):
\begin{gather*}
u_{1}^\sigma(\phi) = t_{\alpha_1}^\sigma(\phi)\qquad\qquad
\tau_{1}^\sigma(\phi) = \theta_{\alpha_1}^\sigma(\phi)\\
u_{i+1}^\sigma(\phi) = \ite{\eq{to(\phi)}{\alpha_{i+1}}}{t_{\alpha_{i+1}}^\sigma(\phi)}{u_{i}^\sigma(\phi)}\\
\tau_{i+1}^\sigma(\phi) = \ite{\eq{to(\phi)}{\alpha_{i+1}}}{\theta_{\alpha_{i+1}}^\sigma(\phi)}{\tau_{i}^\sigma(\phi)}
\end{gather*}
 where  $\ite{b}{\theta_1}{\theta_2}$ denotes the substitution $\theta$ defined by: for every variable $x$, $x\theta =\ite{b}{x\theta_1}{x\theta_2}$.

We then let $t^\sigma_S(\phi) = u_r^\sigma(\phi)$ and  $\theta_S^\sigma(\phi) = \tau_r^\sigma(\phi)$.

\begin{example}

Let us return to the example \ref{example:kcl}.

 Each tag $T_{A_i}$ has an identifier $A_i$ and a key $\key_{A_i}$. In the \kcl protocol the $\textsc{TagInit}_i$ call is
useless because the tag has only one message to send in a round of the protocol ($\textsc{TagInit}_i$ is used to tell a tag to stop the current round of the protocol and to start a new one). Since we consider a finite number $n$ of interface calls, there is at most $n$ sessions executed in parallel. We will use a variable $\textsf{nb}$ to store the index of the next session to start (initialized to $1$), and variables $c_0^1,\dots,c_0^n$ and $c_1^1,\dots,c_1^n$ (initialized to $0$) where $c_0^i$ and $c_1^i$ store respectively the first challenge and the second challenge of session $i$. 
\begin{itemize}
\item $t_{\textsc{SetKey}_i}^\sigma(\phi) = \pair{\sigma(x_{\key_i})}{\sigma(x_{Id_i})}$: the data of the tag $i$ are disclosed.
\item $\theta_{\textsc{SetKey}_i}^\sigma(\phi) = \{x_{\key_i} \mapsto g_{\textsc{key}_i}(\phi), x_{Id_i} \mapsto g_{\textsc{id}_i}(\phi)\}$: the key and id of the tag $i$ are
set to values chosen by the attacker ($g_{\textsc{key}_i}, g_{\textsc{id}_i}\in {\cal G}$).
\item $t_{\textsc{TagMsg}_i}^\sigma(\phi) = $\[\pair{\sigma(x_{Id_i}) \oplus 
\nonce_T
}{\nonce_T \oplus \hash(g_{\textsc{TMsg}_i}(\phi),\sigma(x_{\key_i}))}\]
The reply of the tag $i$ follows the protocol, according to its local store. $g_{\textsc{TMsg}_i}(\phi)$ is the message, forged
by the adversary, replacing the expected name $\nonce_{\textsc{R}}$; $g_{\textsc{TMsg}_i}\in {\cal G}$.
\item $\theta_{\textsc{TagMsg}_i}^\sigma(\phi) = \epsilon$: there is no update in this case (nothing is stored for further verifications in this particular protocol)
\item $t_{\textsc{ReaderInit}}^\sigma(\phi) = \pair{\sigma(\textsf{nb})}{\nonce_R}$: when starting a new session, the reader sends a new challenge $\nonce_R$
\item $\theta_{\textsc{ReaderInit}}^\sigma(\phi) =$
\[ 
  \left\{
    \begin{array}{l}
      \textsf{nb} \mapsto (\sigma(\textsf{nb}) + 1)\\ 
      c^j \mapsto \ite{\eq{\sigma(\textsf{nb})}{j}}{\nonce_R}{\sigma(c^j)} \,|\, 1 \le j \le n
    \end{array}
  \right.
\]
The local memory of the reader is updated.
\end{itemize}

\end{example}

We now express the Privacy game as a set of equivalence properties. After the first phase, once the attacker has chosen the 
corrupted tags, we rename the tags in such a way that the challenged tags are $T_{n-1}$ and $T_n$. In the definition below, $p$ is the number of interactions of the adversary during the first phase and $q$ is the number of interactions of the adversary during the second phase.

\begin{definition}[$m$-Bounded Session Privacy] 
\label{def:bsp}
Given a (computational) interpretation $I$ of the function symbols in $\cal F$,
a protocol satisfies \emph{$m$-Bounded Session Privacy} if for every $p,q$ such that $p+q = m$ and for every computational model ${\cal M}_c^I$ that extends
$I$, we have:
\begin{multline*}
 {\cal M}_c^I \models (t^{\sigma_i}(\phi_i))_{ i \le m},g_{\text{guess}}(\phi_{m+1}) \\
\sim (t^{\tilde \sigma_i}(\tilde \phi_i))_{ i \le m},g_{\text{guess}}(\tilde\phi_{m+1})
\end{multline*}
where $\phi_1=\epsilon$, $\sigma_1 = \sigma_{\text{init}}$ and for all $1 < i \le m$:
\begin{itemize}
\item $\phi_{i+1} = 
\begin{cases}
\phi_i,t_{\Gamma_n}^{\sigma_i}(\phi_i) \text{ if }i \le p\\
\phi_i,t_{\Gamma_{n-1} \backslash \{ \textsc{SetKey}_{n-1}\}}^{\sigma_i}(\phi_i) \text{ if } i \ge p
\end{cases}$
\item $\sigma_{i+1} = 
\begin{cases}
\sigma_i\,\theta^{\sigma_i}_{\Gamma_n}(\phi_i) \hfill(\text{ if }i \le p)\\
\sigma_i\,\{x_{\key_{n-1}} \mapsto \sigma_i(x_{\key_{n-1}})\}\,\theta^{\sigma_i}_{\Gamma_{n-1} \backslash \{ \textsc{SetKey}_{n-1}\}}(\phi_i) \\\hfill(\text{ if }i = p+1)\\
\sigma_i\,\theta^{\sigma_i}_{\Gamma_{n-1} \backslash \{ \textsc{SetKey}_{n-1}\}}(\phi_i) \hfill(\text{ if }i > p+1)
\end{cases}$
\end{itemize}
$\tilde \phi_i$ and $\tilde\sigma_i$ are defined similarly, except that the chosen tag is $T_n$ and not $T_{n-1}$ (that is the key substitution used is $\{x_{\key_{n-1}} \mapsto \tilde\sigma_i(x_{\key_{n}})\}$ instead of $\{x_{\key_{n-1}} \mapsto \sigma_i(x_{\key_{n-1}})\}$ in the case $i=p+1$). $g_{\text{guess}}\in {\cal G}$ is the attacker's function guessing with which tag the interaction took place.
\end{definition}

\begin{theorem}
Given a computational interpretation of function symbols in $\cal F$, a protocol $P$ satisfies $m$-Bounded Session Privacy iff it satisfies $m$-Privacy.
\end{theorem}

\begin{IEEEproof}[Proof sketch]
If there is an attacker on $m$-privacy, we can construct an interpretation of the attacker's function symbols that yields a counter-model of the equivalence property. Conversely, if there is a counter-model we can build an attacker that performs the actions and computations specified by the interpretations of these function symbols. These constructions correspond basically to the proofs in \cite{bana14ccs}.
\end{IEEEproof}

\subsection{Fixed Trace Privacy}

The $m$-Bounded Session Privacy definition is a bit cumbersome, since the terms gather together all possible trace interleavings
of the protocol with $n$ tags and $m$ calls to the interfaces. It is easier to use the following definition, that we call $m$-Fixed Trace Privacy, that considers equivalence formulas between executions of the protocol with a  fixed sequence of actions chosen by the adversary. Basically we split a big equivalence into an exponential number  (in $m$) of smaller formulas.
\begin{definition}[$m$-Fixed Trace Privacy] 
\label{def:ftp}
Given an interpretation $I$ of function symbols in $\cal F$, a protocol satisfies $m$-Fixed Trace Privacy if for all $p,q$ such that $p+q = m$, for all $(\alpha_i)_{1 \le i \le m} \in (\Gamma_n)^p \times (\Gamma_{n-1} \backslash \textsc{SetKey}_{n-1})^q$, for all computational models
$\cmodel$ that extend $I$, we have:
\begin{multline*}
 \cmodel \models (t^{\sigma_i}_{\alpha_i}(\phi_i))_{ i \le m},g_{\text{guess}}(\phi_{m+1}) \\
\sim (t^{\tilde \sigma_i}_{\alpha_i}(\tilde \phi_i))_{ i \le m},g_{\text{guess}}(\tilde\phi_{m+1})
\end{multline*}
where $\phi_1=\epsilon$, $\sigma_1 = \sigma_{\text{init}}$ and for all $1 < i \le m$:
\begin{itemize}
\item $\phi_{i+1} = \phi_i,t^{\sigma_i}_{\alpha_i}(\phi_i)$
\item $\tilde \phi_{i+1} = \tilde \phi_i,t^{\tilde \sigma_i}_{\alpha_i}(\tilde \phi_i)$
\item $\sigma_{i+1} = 
\begin{cases}
\sigma_i\,\theta^{\sigma_i}_{\alpha_i}(\phi_i) \text{ if }i \ne p+1\\
\sigma_i\,\{x_{\key_{n-1}} \mapsto \sigma_i(x_{\key_{n-1}})\}\,\theta^{\sigma_i}_{\alpha_i}(\phi_i) \text{ if }i = p+1
\end{cases}$
\item $\tilde\sigma_{i+1} = 
\begin{cases}
\tilde\sigma_i\,\theta^{\tilde\sigma_i}_{\alpha_i}(\tilde\phi_i) \text{ if }i \ne p+1\\
\tilde\sigma_i\,\{x_{\key_{n-1}} \mapsto \tilde\sigma_i(x_{\key_{n}})\}\,\theta^{\tilde\sigma_i}_{\alpha_i}(\tilde\phi_i) \text{ if }i = p+1
\end{cases}$
\end{itemize}
\end{definition}

\begin{proposition}\label{proposition:trace}
$m$-Fixed Trace Privacy is equivalent to $m$-Bounded Session Privacy.
\end{proposition}

\begin{IEEEproof}
We start by showing that $m$-Bounded Session Privacy implies $m$-Fixed Trace Privacy by contraposition:  let $p,q$ be two integers such that $p+q = m$, $(\alpha_i)_{1 \le i \le m} \in (\Gamma_n)^p \times (\Gamma_{n-1} \backslash \textsc{SetKey}_{n-1})^q$ and $\cmodel$ be a computational model such that:
\begin{multline*}
 \cmodel \models (t^{\sigma_i}_{\alpha_i}(\phi_i))_{i \le m},g_{\text{guess}}(\phi_{m+1}) \\
\not \sim (t^{\tilde \sigma_i}_{\alpha_i}(\tilde \phi_i))_{ i \le m},g_{\text{guess}}(\tilde\phi_{m+1})
\end{multline*}
Let $\mathcal{A}$ be an adversary with a non negligible probability of distinguishing between the two distributions (interpreted in \cmodel).  We define the model $\cmodel'$ to be $\cmodel$ with a signature extended by the function symbols $\{ to \} \cup \Gamma_n$: for all $\alpha \in \Gamma_n$, $\llbracket \alpha \rrbracket_{\cmodel'}$ is interpreted by a machine returning $\alpha$ on all input, and the function symbol $to$ is interpreted by a machine satisfying:
\[
  \llbracket to(\phi_i) \rrbracket_{\cmodel'} = \llbracket to(\tilde \phi_i)\rrbracket_{\cmodel'} = \llbracket \alpha_i \rrbracket_{\cmodel'}
\]
 This only depends on the size of $\phi_i$ and $\tilde \phi_i$, and is therefore possible to do with a polynomial time probabilistic Turing Machine, hence $\cmodel'$ is a computational model. Moreover by construction $\mathcal{A}$ has a non negligible probability of breaking $m$-Bounded Session Privacy in $\cmodel'$, which concludes the proof. 

Conversely we show that $m$-Fixed Trace Privacy implies $m$-Bounded Session Privacy by contraposition: using notation of Definition~\ref{def:bsp}, let $p,q$ be such that $p+q = m$ and assume a model $\cmodel$ and an adversary $\mathcal{A}$ with a non negligible probability of distinguishing between $ \llbracket\phi_{m+1}\rrbracket$ and $ \llbracket\tilde\phi_{m+1}\rrbracket$ (we remove $\llbracket g_{guess}(\phi_{m+1}) \rrbracket)$ since $\mathcal{A}$ can compute it himself). We let $S = (\Gamma_n)^p \times (\Gamma_{n-1} \backslash \textsc{SetKey}_{n-1})^q$ and for all $\vec \alpha \in S$, $E_{\vec \alpha}$ is the event $\bigwedge_i \llbracket to(\phi_i)\rrbracket_{\eta} = \alpha_i$. We then have:
\begin{align*}
&\Pr_\rho(\mathcal{A}(\llbracket\phi_{m+1}\rrbracket_{\eta,\rho}) = 1) \\
&=\! \sum_{\vec \alpha \in S} \Pr_\rho( \mathcal{A}(\llbracket\phi_{m+1}\rrbracket_{\eta,\rho}) = 1 \mid E_{\vec\alpha} ) \times \Pr_\rho( E_{\vec\alpha})\\
&=\! \sum_{\vec \alpha \in S} \Pr_\rho( \mathcal{A}(\llbracket(t^{\sigma_i}_{\alpha_i}(\psi^{\vec \alpha}_i))_{ i \le m}\rrbracket_{\eta,\rho}) = 1 ) \times \Pr_\rho( E_{\vec\alpha})
\end{align*}
where $\psi^{\vec \alpha}_i$ denotes the $i$-th frame obtained for the fixed trace $\vec \alpha$ (it is the $\phi_i$ from Definition~\ref{def:ftp} renamed to avoid notations clash). Similarly we let $E'_{\vec \alpha}$ is the event $\bigwedge_i \llbracket to(\tilde \phi_i)\rrbracket_{\eta} = \alpha_i$ and we have:
\begin{align*}
&\Pr_\rho(\mathcal{A}(\llbracket\tilde \phi_{m+1}\rrbracket_{\eta,\rho}) = 1) \\
&=\! \sum_{\vec\alpha\in S} \Pr_\rho( \mathcal{A}(\llbracket(t^{\tilde \sigma_i}_{\alpha_i}(\tilde \psi^{\vec \alpha}_i))_{ i \le m}\rrbracket_{\eta,\rho}) = 1 ) \times \Pr_\rho( E'_{\vec \alpha})
\end{align*}

Since $\Pr_\rho(E_{\vec \alpha}) \le 1$, $\Pr_\rho(E'_{\vec \alpha}) \le 1$ and bounding the absolute value of the sum by the sum of the absolute values we get that:
\[
    \big|\Pr_\rho(\mathcal{A}(\llbracket \phi_{m+1}\rrbracket_{\eta,\rho}) = 1) - \Pr_\rho(\mathcal{A}(\llbracket\tilde \phi_{m+1}\rrbracket_{\eta,\rho}) = 1)\big|
\]
is upper bounded by:
\begin{multline*}
 \sum_{\vec\alpha\in S} \big|\Pr_\rho( \mathcal{A}(\llbracket(t^{ \sigma_i}_{\alpha_i}( \psi^{\vec \alpha}_i))_{ i \le m}\rrbracket_{\eta,\rho}) = 1 ) -\\
    \Pr_\rho( \mathcal{A}(\llbracket(t^{\tilde \sigma_i}_{\alpha_i}(\tilde \psi^{\vec \alpha}_i))_{ i \le m}\rrbracket_{\eta,\rho}) = 1 ) \big|
\end{multline*}
Since $\mathcal{A}$ distinguishes $ \llbracket\phi_{m+1}\rrbracket$ and $ \llbracket\tilde\phi_{m+1}\rrbracket$ with non negligible probability, and since $S$ is finite we know that there exists $\vec \alpha$ such that:
\begin{multline*}
\big|\Pr_\rho( \mathcal{A}(\llbracket(t^{ \sigma_i}_{\alpha_i}( \psi^{\vec \alpha}_i))_{ i \le m}\rrbracket_{\eta,\rho}) = 1 ) -\\
    \Pr_\rho( \mathcal{A}(\llbracket(t^{\tilde \sigma_i}_{\alpha_i}(\tilde \psi^{\vec \alpha}_i))_{ i \le m}\rrbracket_{\eta,\rho}) = 1 ) \big|
\end{multline*}
is non negligible. Therefore $m$-Fixed Trace Privacy does not hold, which concludes this proof.


\end{IEEEproof}

As explained in \cite{bana14ccs}, indistinguishability properties can be expressed in the logic for
arbitrary \emph{determinate} protocols. For such protocols, observational equivalence and trace
equivalence coincide \cite{cortier09csf}. The above proposition is a similar result in the computational model. It
can be extended to other equivalence properties, as long as $m$ does not depend on the security parameter.



\section{Two \rfid protocols}
\label{section:protocols}
 We are now going to describe  the $\lak$ and $\kcl$ \rfid protocols, as well,  attacks, patches and formal (computational)
security proofs of the fixed versions. 

We first consider that names are randomly generated numbers, even though, because of the limited computing capabilities of the tags,
they have to be implemented using a Cryptographic Pseudo Random Number Generator (\cprng). This issue will be discussed in the section \ref{section:prng}: we
will show that we can always safely abstract the pseudo random numbers as random numbers, provided that a \cprng is used and the
random seed is never used for other purposes.

\subsection{A known attack on \kcl}
Let us return to the simple example described in Figure~\ref{figure:kcl}:
\[
  \begin{array}{l}
    \begin{array}{lcl}
      R&:& \nonce_{R} \rsample\\
      T_{\textsf{A}}&:& \nonce_{T} \rsample\\
    \end{array} \\[2em]
    \begin{array}{lcl}
      1: R \longrightarrow T_{\textsf{A}} &:& \nonce_R\\
      2: T_{\textsf{A}} \longrightarrow R &:& \pair{\textsf{A} \oplus \nonce_T}{\nonce_T \oplus \hash(\nonce_R,\key_\textsf{A})}
    \end{array}
  \end{array}
\]

As reported in \cite{van2008attacks}, there is a simple attack that we depict in Figure~\ref{figure:kcl-attack}. In this attack the tag is challenged twice with the same name:
observing the exchanges between the tag and the reader, the adversary can replay the name. Finally the adversary checks if he is
talking with the same tag by xoring the two components of the message sent by the second tag, and verifies whether the result is the same as what he
obtained with the same operation in the first session.

\begin{figure*}
\noindent\framebox{\parbox{0.98\linewidth}{
\[
  \begin{array}{lcl|lcl}
    1: R \longrightarrow T_{\textsf{A}} &:& \nonce_R & R \longrightarrow T_{\textsf{A}} &:& \nonce_R\\
    2: T_{\textsf{A}} \longrightarrow R &:& \pair{\textsf{T}_{\textsf{A}} \oplus \nonce_T}{\nonce_T \oplus \hash(\nonce_R,\key_\textsf{A})}& T_{\textsf{A}} \longrightarrow R &:& \pair{\textsf{T}_{\textsf{A}} \oplus \nonce_T}{\nonce_T \oplus \hash(\nonce_R,\key_\textsf{A})}\\[1em]
    3: E \longrightarrow T_{\textsf{A}}&:& \nonce_R & E \longrightarrow T_{\textsf{B}}&:& \nonce_R\\
    4: T_{\textsf{A}} \longrightarrow R &:& \pair{\textsf{T}_{\textsf{A}} \oplus \nonce'_T}{\nonce'_T \oplus \hash(\nonce_R,\key_\textsf{A})}& T_{\textsf{B}} \longrightarrow R &:& \pair{\textsf{T}_{\textsf{B}} \oplus \nonce'_T}{\nonce'_T \oplus \hash(\nonce_R,\key_\textsf{B})}\\
  \end{array}
\]
}}
\caption{Attack against the original \kcl protocol}
\label{figure:kcl-attack}
\end{figure*}

In the left execution, the xor of the two part of the tag answers will be the same:
\begin{align*}
\textsf{T}_{\textsf{A}} \oplus \nonce_T \oplus \nonce_T \oplus \hash(\nonce_R,\key_\textsf{A}) &= \textsf{T}_{\textsf{A}} \oplus \nonce'_T \oplus \nonce'_T \oplus \hash(\nonce_R,\key_\textsf{A})\\
&= \textsf{T}_{\textsf{A}}  \oplus \hash(\nonce_R,\key_\textsf{A})
\end{align*}
Whereas in the right execution we will obtain two values $\textsf{T}_{\textsf{A}}  \oplus \hash(\nonce_R,\key_\textsf{A})$ and $\textsf{T}_{\textsf{B}}  \oplus \hash(\nonce_R,\key_\textsf{B})$ which will be different with high probability. 

\subsection{\kclp, a revised version of  \kcl}
\label{section:kcl-verif}

We propose a simple correction to the \kcl protocol: we replace the first occurrence of the name $\nonce_T$ with its hash, breaking the algebraic property that was used in the attack. This protocol is depicted in Figure~\ref{fig:kclp}, and to our knowledge there exists no formal study of this revised version.
\begin{figure}[H]
\[
  \begin{array}{l}
    \begin{array}{lcl}
      R&:& \nonce_{R} \rsample\\
      T&:& \nonce_{T} \rsample
    \end{array}\\[2em]
    \begin{array}{lcl}
      1: R \longrightarrow T_{\textsf{A}} &:& \nonce_R\\
      2: T_{\textsf{A}} \longrightarrow R &:& \pair{\textsf{A} \oplus \hash(\nonce_T,\key_{\textsf{A}})}{\nonce_T \oplus \hash(\nonce_R,\key_{\textsf{A}})}
    \end{array} 
  \end{array} 
\]
\caption{\label{fig:kclp} The \kclp protocol}
\end{figure}

We are now going to illustrate our method  by showing that the \kclp protocol verifies $m$-Fixed Trace Privacy with two tags \textsf{A} and \textsf{B}. Assuming collision resistance only, there is actually an attack on the protocol \kclp (exactly the same attack as the one described in Section~\ref{section:lak}). We therefore assume the PRF property.
\begin{theorem}[Unlinkability for an arbitrary number of rounds]
Assuming PRF for the keyed hash function, 
the \kclp protocol verifies $m$-Fixed Trace Privacy for two agents and all $m$.
\end{theorem}

In the proof below, the primed version of a term $t$ is the term $t$, in which  the names $\nonce_1,\dots,\nonce_l$ appearing in $t$ have been replaced by $\nonce'_1,\dots,\nonce'_l$. We will use $t^\textsf{\textsf{Id}}_{\phi}$ (where $\textsf{Id} = \textsf{A}$ or $\textsf{B}$) to denote the response of the tag $T_{\textsf{Id}}$ to a challenge:
\[t^\textsf{\textsf{Id}}_{\phi} = \pair{\textsf{Id} \oplus \hash(\nonce_T,\key_\textsf{Id})}{\nonce_T \oplus \hash(g(\phi),\key_\textsf{Id})}\]
Since the axioms are Horn clauses, we can view them as inference rules, in which the premises of the inferences are the negative literals of the clause. This is easier to display and read. 

\begin{IEEEproof}
We prove this by induction on $m$ (induction is outside our logic). Let $\phi,\tilde \phi$ be two sequences of terms from the $m$-Fixed Trace Privacy definition. By induction hypothesis, we assume that we have a derivation of $\phi \sim \tilde \phi$ (in the base case, this is the reflexivity of $\sim$). We  consider two cases.

 If the adversary decides to start a new session with the reader, we need to show that $\phi, \nonce_R \sim \tilde \phi, \nonce_R$ where $\nonce_R$ is fresh in $\phi,\tilde \phi$. This case is easy, we only need to apply the $FreshNonce$ axiom:
\begin{prooftree}
\ax{\phi \sim \tilde \phi}
\uinf{\phi, \nonce_R \sim \tilde \phi, \nonce_R}{FreshNonce}
\end{prooftree}

Otherwise, the adversary decides to interact with the tags, e.g. $\textsf{A}$ on the left and $\textsf{B}$ on the right (the other cases are identical). We want to show that \(\phi,t^{\textsf{A}}_{\phi} \sim \tilde \phi,\tilde t^{\textsf{B}}_{\tilde \phi}\). We let $\nonce$ be a fresh name and $\psi,\tilde \psi$ be defined by:
\begin{gather*}
\psi \equiv \phi, \nonce_T \oplus \hash(g(\phi),\key_\textsf{A})\\
\tilde \psi \equiv  \tilde \phi,  \nonce_T \oplus \hash(g(\tilde \phi),\key_\textsf{B})
\end{gather*}
We start (from the root) our proof by applying the $FA$  axiom (breaking the pair) and then to introduce an intermediate term $\textsf{A}\oplus \nonce$ since, intuitively, $\hash(\nonce_T,\key_\textsf{A})$ (resp. $\hash(\nonce_T,\key_\textsf{B})$) should be indistinguishable from a random number.

\begin{prooftree}
\ax{\psi, \hash(\nonce_T,\key_\textsf{A}) \sim \psi, \nonce}
\uinf{\psi,\textsf{A} , \hash(\nonce_T,\key_\textsf{A}) \sim \psi,\textsf{A}, \nonce}{FA}
\uinf{\psi,\textsf{A} \oplus \hash(\nonce_T,\key_\textsf{A}) \sim \psi,\textsf{A} \oplus \nonce}{FA}
\ax{P_1}
\binf{\psi,\textsf{A} \oplus \hash(\nonce_T,\key_\textsf{A})
\sim  \tilde \psi,\textsf{B} \oplus \hash(\nonce_T,\key_\textsf{B})}{Trans}
\uinf{\phi, t^A_{\phi}
\sim \tilde \phi,  t^B_{\tilde \phi}}{FA}
\end{prooftree}
where $P_1$ is a derivation of $  \psi,\textsf{A} \oplus \nonce \sim \tilde \psi,\textsf{B} \oplus \hash(\nonce_T,\key_\textsf{B})$.

\paragraph{Left Derivation}
We have to find first a  a derivation of \(\psi, \hash(\nonce_T,\key_\textsf{A}) \sim \psi, \nonce\).
The ultimate goal is to apply the $PRF$ axioms. For that, we need to introduce, on both sides of the $\sim$ predicate, equality tests between the last message hashed under key $\key_{\textsf{A}}$ (i.e. $\nonce_T$), and all the previous hashed messages under key $\key_{\textsf{A}}$. We let $m_1,\dots,m_l$ be the set of messages hashed with $\key_{\textsf{A}}$ in $\phi$. We know that these messages are either names $\nonce_T'$, or of the form $g(\phi')$ where $\phi'$ is a strict prefix of $\phi$. 

Let $\alpha = \hash(\nonce_T,\key_\textsf{A})$, $\beta = \nonce$. For all $ 1 \le i \le l$ we let $e_i \equiv \eq{\nonce_T}{m_i}$, and $s^x$ be the term:
\begin{gather*}
    \ite{e_1}{x}{\phantom{x}\\
      \ite{e_2}{x}{\phantom{x}\\
        \rotatebox{90}{\dots}\\
        \ite{e_l}{x}{x}
      }
    }
\end{gather*}
We observe that, for every term $u$, $u = s^u$ is derivable from the equality axioms.
 We are now going to use the $CS$ axiom to split the proof. To do so we introduce for every $1 \le i \le l$ the term $u_i^x$:
 \begin{gather*}
   \ite{e_1}{0}{\\
     \rotatebox{90}{\dots}\\
     \ite{e_{i-1}}{0}{\\
       \ite{e_{i}}{x}{0}
    }
  }
\end{gather*}
And the term $u_{l+1}^x$:
\[\ite{e_1}{0}{\dots\ite{e_l}{0}{x}}\]
By repeatedly applying the $CS$ axiom we obtain:
\vspace{-1.4em}
\begin{prooftree}
\ax{\forall i \in \{1,\dots,l+1\},\;\psi,e_1,\dots,e_l,u_i^\alpha \sim \tilde \psi,e_1,\dots,e_l,u_i^\beta}
\uinf{\psi, s^{\hash(\nonce_T,\key_\textsf{A})} \sim  \psi, s^{\nonce}}{CS}
\uinf{\psi, \hash(\nonce_T,\key_\textsf{A}) \sim  \psi, \nonce}{Congr}
\end{prooftree}
First note that, using the $EqIndep$ axiom, we derive,  for every $1 \le i \le l$, $e_i = \false$. This allows us to deal with cases $1$ to $l$, since this implies that $u_i^\alpha = u_i^\alpha = 0$ is derivable. Therefore we have for all $i \in \{1,\dots,l\}$:
\begin{prooftree}
\ax{}
\uinf{\psi,\false,\dots,\false,0 \sim \psi,\false,\dots,\false,0}{Refl}
\uinf{\psi,e_1,\dots,e_l,u_i^\alpha \sim \psi,e_1,\dots,e_l,u_i^\beta}{Congr}
\end{prooftree}

Consider now the case $i=l+1$. The conditions on the occurrences of $\hash$ and $\key_{\textsf{A}}$ are satisfied,
thanks to the choice of $e_1,\dots,e_l$. We may use the $PRF_l$ axiom: 
\vspace{-1.4em}
\begin{prooftree}
\ax{}
\uinf{\psi,u_1^\alpha \sim \psi,u_1^\beta}{PRF}
\uinf{\psi,\false,\dots,\false,u_{l+1}^\alpha \sim \psi,\false,\dots,\false,u_{l+1}^\beta}{FA^{(l)}}
\uinf{\psi,e_1,\dots,e_l,u_1^\alpha \sim \psi,e_1,\dots,e_l,u_1^\beta}{Congr}
\end{prooftree}

\paragraph{Right Derivation ($P_1$)} Now, we have to derive \(\psi,\textsf{A} \oplus \nonce \sim \tilde \psi,\textsf{B} \oplus \hash(\nonce_T,\key_\textsf{B})\). We start by replacing $\textsf{A}$ with $\textsf{B}$, splitting again the proof in two subgoals:
\vspace{-0.5em}
\begin{prooftree}\small
\ax{ \psi,\textsf{A} \oplus \nonce \sim \tilde \psi,\textsf{B} \oplus \nonce}
\ax{ \tilde \psi,\textsf{B} \oplus \nonce \sim \tilde \psi,\textsf{B} \oplus \hash(\nonce_T,\key_\textsf{B})}
\binf{ \psi,\textsf{A} \oplus \nonce \sim \tilde \psi,\textsf{B} \oplus \hash(\nonce_T,\key_\textsf{B})}{Trans}
\end{prooftree}

For the right part, we first decompose the goal:
\vspace{-0.5em}
\begin{prooftree}
\ax{ \tilde \psi, \hash(\nonce_T,\key_\textsf{B}) \sim \tilde \psi, \nonce }
\uinf{\tilde \psi, \nonce \sim \tilde \psi, \hash(\nonce_T,\key_\textsf{B})}{Sym}
\uinf{\tilde \psi,\textsf{B}, \nonce \sim \tilde \psi,\textsf{B}, \hash(\nonce_T,\key_\textsf{B})}{FA}
\uinf{\tilde \psi,\textsf{B} \oplus \nonce \sim \tilde \psi,\textsf{B} \oplus \hash(\nonce_T,\key_\textsf{B})}{FA}
\end{prooftree}
Then, the derivation of $ \tilde \psi, \hash(\nonce_T,\key_\textsf{B}) \sim \tilde \psi, \nonce $ is similar to the derivation of $\psi, \hash(\nonce_T,\key_\textsf{A}) \sim \psi, \nonce$.

For the left part, we start by using the axiom on $\xor$:
\begin{prooftree}
  \ax{\psi\sim \tilde\psi}
  \uinf{ \psi,\textsf{A} \oplus \nonce \sim \tilde \psi,\textsf{B} \oplus \nonce}{Indep}
\end{prooftree}

It only remains to show that $ \psi \sim \tilde \psi$. We do this using the $Trans$ and $Indep$ axioms:
\begin{align*}
\lefteqn{\overbrace{\phi, \nonce_T \oplus \hash(g(\phi),\key_\textsf{A}) \sim \phi, \nonce_T}^{\textsf{LSim}}}
\lefteqn{\phi, \nonce_T \oplus \hash(g(\phi),\key_\textsf{A}) \sim \underbrace{\phi, \nonce_T \sim \tilde \phi, \nonce_T}_{\textsf{MSim}} \sim }
\phi, \nonce_T \oplus \hash(g(\phi),\key_\textsf{A}) \sim \phi, \nonce_T \sim  \overbrace{\tilde \phi, \nonce_T \sim  \phi, \nonce_T \oplus \hash(g(  \phi),\key_\textsf{B})}^{\textsf{RSim}}
\end{align*}

\begin{prooftree}
\ax{}
\uinf{\phi \sim \phi}{Refl}
\uinf{\textsf{LSim}}{Indep}
\ax{\phi \sim \tilde \phi}
\uinf{\textsf{MSim}}{Indep}
\ax{} 
\uinf{\phi \sim \phi}{Refl}
\uinf{\textsf{RSim}}{Indep}
\tinf{\phi, \nonce_T \oplus \hash(g(\phi),\key_\textsf{A})
\sim  \phi, \nonce_T \oplus \hash(g(  \phi),\key_\textsf{B})}{Trans}
\end{prooftree}
Which concludes the inductive step proof.
\end{IEEEproof}



In this result we consider only two tags, for simplicity reasons. In particular they cannot be corrupted tags. Our framework is expressive enough for multiple tags, including corrupted one, though we did not complete the proof in that case.

\subsection{The \lak  protocol}
\paragraph{Description} We describe in Figure~\ref{figure:lak} 
the original protocol from \cite{lee2006rfid}. As we mentioned before, the protocol we consider 
is a simplified version of the \lak protocol, without the key server. In the \lak protocol, the reader shares a private key $\key_{\textsf{A}}$ with each of its tags $T_{\textsf{A}}$. $\ow$ is an hash function. This is a stateful protocol: the key is updated after each successful completion of the protocol, and the reader keeps in  $\key^0_{\textsf{A}}$ the previous value of the key. This value is used as a backup in case $T_{\textsf{A}}$ has not completed the protocol (for example because the last message was lost) and therefore not updated its version of the key. The protocol allows to recover from such a desynchronization: the reader $R$ can use the previous version of $\key_{\textsf{A}}$ at the next round (which is the version used by $T_{\textsf{A}}$) and finish the protocol.

\begin{figure}
\[
  \begin{array}{l}
    \begin{array}{lcl}
      R&:& \nonce_{R} \rsample\\
      T_{\textsf{A}}&:& \nonce_{T} \rsample\\
    \end{array}\\[2em]
    \begin{array}{lcl}
    1: R \longrightarrow T_\textsf{A} &:& \nonce_R\\
    2: T_\textsf{A} \longrightarrow R &:& \pair{\nonce_T}{\ow(\nonce_R \oplus \nonce_T \oplus \key_\textsf{A})}\\
    3: R \longrightarrow T_\textsf{A} &:& \ow\big((\ow(\nonce_R \oplus \nonce_T \oplus \key_\textsf{A})\oplus\nonce_R \oplus \key_\textsf{A}\big)\\
  \end{array}\\[2em]
  \begin{array}{lcl}
    R&:&\key_\textsf{A} = \ow(\key_\textsf{A})\;,\; \key_\textsf{A}^0 = \key_\textsf{A}\\
    T_\textsf{A}&:&\key_\textsf{A} = \ow(\key_\textsf{A})
  \end{array}
\end{array}
\]
\caption{The \textsc{lak} protocol}
\label{figure:lak}
\end{figure}

The protocol is supposed to achieve mutual authentication and unlinkability. Even though such properties 
can be defined in various ways, we recall below a known attack against the \lak protocol, which will force us to 
modify  it. 

\paragraph{An attack on \lak}
An attack on authentication is described in \cite{van2008attacks} and sketched below:
\[\begin{array}{ll}
1:& R \longrightarrow T_\textsf{A} : \nonce_R\\
2:& T_\textsf{A} \longrightarrow E : \pair{\nonce_T}{\ow(\nonce_R \oplus \nonce_T \oplus \key_\textsf{A})}\\
&\null\\
3:& R \longrightarrow E : \nonce_R'\\
4:& E \longrightarrow R : \pair{ \nonce_R' \oplus \nonce_R \oplus \nonce_T}{\ow(\nonce_R \oplus \nonce_T \oplus \key_\textsf{A})}\\
5:& R \longrightarrow E : \ow\big((\ow(\nonce_R \oplus \nonce_T \oplus \key_\textsf{A})\oplus\nonce_R' \oplus \key_\textsf{A}\big)
\end{array}\]

In this attack, the adversary simply observes the beginning of an honest execution of the protocol (without completing the protocol, so that the reader and the tag do not update the key) between a tag $A$ and the reader.
The adversary obtains $\ow(\nonce_R \oplus \nonce_T \oplus \key_\textsf{A})$ and the names $\nonce_R,\nonce_T$  . 
He then interacts with the reader to get a new name $\nonce_R'$ and impersonates the tag $A$ by choosing 
the returned tag $\nonce_T'$ such that $\nonce_R' \oplus \nonce_T' = \nonce_R \oplus \nonce_T$.

\subsection{A stateless revised version of \lak}
\label{section:lak}
In \cite{hirschi16sp}, the authors consider a corrected (and stateless) version of the protocol,
which they proved secure. This version of the protocol is described below:
\[
  \begin{array}{l}
    \begin{array}{lcl}
      R&:& \nonce_{R} \rsample\\
      T_\textsf{A}&:& \nonce_{T} \rsample\\ \\
    \end{array}\\
    \begin{array}{lcl}
      1: R \longrightarrow T_\textsf{A} &:& \nonce_R\\
      2: T_\textsf{A} \longrightarrow R &:& \pair{\nonce_T}{\ow(\langle \nonce_R, \nonce_T, \key_\textsf{A}\rangle)}\\
      3: R \longrightarrow T_\textsf{A} &:& \ow\big(\langle(\ow(\langle\nonce_R, \nonce_T, \key_\textsf{A}\rangle), \nonce_R , \key_\textsf{A} \rangle\big)\\
    \end{array}
  \end{array} 
\]

This new version avoids the previous attack, which relied on the algebraic properties of exclusive-or.
Formally, the protocol is described in the applied pi-calculus in \cite{hirschi16sp}, in which they prove the strong unlinkability property of
\cite{arapinis10} in the Dolev-Yao model for an unbounded number of session.

\paragraph{Attack against stateless \lak}
Since the stateless version of \lak was proved in the symbolic model, no computational security assumptions were made on $\ow$. We show in Figure~\ref{fig:owattack} 
 that choosing  $\ow$ to be a one-way cryptographic hash function (OW-CPA and Strongly Collision Resistant for example) is not enough to guarantee unlinkability.
\begin{figure*}
  \noindent\framebox{\parbox{0.98\linewidth}{
      \[
        \begin{array}{lcl|lcl}
          1: E \longrightarrow T_{\textsf{A}} &:& \nonce_R & E \longrightarrow T_{\textsf{A}} &:& \nonce_R\\
          2: T_{\textsf{A}} \longrightarrow E &:& \pair{\nonce_T}{{\ow(\langle \nonce_R, \nonce_T, \key_\textsf{A}\rangle)}} & T_{\textsf{A}} \longrightarrow E &:& \pair{\nonce_T}{{\ow(\langle \nonce_R, \nonce_T, \key_\textsf{A}\rangle)}}\\
                                              &\null&\\
          3: E \longrightarrow T_{\textsf{A}} &:& \nonce_R' & E \longrightarrow T_{\textsf{B}} &:& \nonce_R'\\
          4: T_{\textsf{A}} \longrightarrow E &:& \pair{\nonce_T'}{{\ow(\langle \nonce_R', \nonce_T', \key_\textsf{A}\rangle)}} & T_{\textsf{B}} \longrightarrow E &:& \pair{\nonce_T'}{{\ow(\langle \nonce_R', \nonce_T', \key_\textsf{B}\rangle)}}
        \end{array}
      \]
    }}
  \caption{\label{fig:owattack}Unlinkability attack in two rounds against the stateless \lak protocol}
\end{figure*}

The attack is quite simple: it suffices that the hash function $\ow$ leaks a few bits of the  hashed message (which is possible for an one-way hash function). This means that, when hashing a message of the form $\langle \nonce_R, \nonce_T, \key\rangle$, the hash function $\ow$ will leak some bits of the agent key $\key$. Since the keys are drawn uniformly at random, there is a non negligible probability for the leaked bits to be different when hashing messages with different keys. In particular an adversary will be be able to distinguish $\ow(\langle \nonce_R, \nonce_T, \key_\textsf{A}\rangle),\ow(\langle \nonce'_R, \nonce'_T, \key_\textsf{A}\rangle)$ from $\ow(\langle \nonce_R, \nonce_T, \key_\textsf{A}\rangle),\ow(\langle \nonce'_R, \nonce'_T, \key_\textsf{B}\rangle)$ with high probability. 

Observe that this attack would still work if we modified the protocol to update the keys after a successful execution of the protocol (in other word, if we consider the original \lak protocol with concatenation instead of xor), because the attacker could start executions of the protocol without finishing them, preventing the keys from being updated.

\begin{remark}
In the original paper introducing \lak \cite{lee2006rfid}, the hash function is described as a one-way cryptographic hash function, which a priori does not prevent the attack described above. However, in the security analysis section, the authors assume the function to be indistinguishable from a random oracle, which prevents the attack. It is actually sufficient to assume
PRF, for which there are effective constructions (subject to hardness assumptions). 
\end{remark}

\subsection{The \lakp protocol} 
We describe here a stateless version of the \lak protocol, that we call \lakp. As in the \lak protocol, the reader shares with each tag a secret key $\key$. We use a keyed-hash function that is assumed to be PRF to prevent the attack depicted in Section~\ref{section:lak}. This protocol uses a function $\combine$ that combines the names. It could be a priori a xor, as in the original protocol, or a pairing, as in the revised version of \cite{hirschi16sp} or something else. We will look for sufficient conditions on this function $\combine$, such that the protocol is secure.
\[\begin{array}{l}
    \begin{array}{lcl}
      R&:& \nonce_{R} \rsample\\ 
      T&:& \nonce_{T} \rsample \\ 
    \end{array}\\[2em]
    \begin{array}{lcl}
      1: R \longrightarrow T &:& \nonce_R\\
      2: T \longrightarrow R &:& \pair{\nonce_T}{\hash(\combine(\nonce_R,\nonce_T),\key_\textsf{A})}\\
      3: R \longrightarrow T &:& \hash\big(\combine(\hash(\combine(\nonce_R,\nonce_T),\key_\textsf{A}),\nonce_R),\key_\textsf{A}\big)
    \end{array}
\end{array}\]

We start by describing two different attacks that rely on some properties of the function $\combine$. In each case, we  give a sufficient condition on $\combine$ that prevents the attack. Next, we  show that these two conditions are sufficient to prove that the \lakp protocol verifies the Bounded Session Privacy property.


\paragraph{First Attack:} The attack depicted 
below is a generalization of the attack from~\cite{van2008attacks}.
It works when
there exists a function $s$ (computable in probabilistic polynomial time) such that the quantity below is not negligible:
\begin{equation}
\Pr\big( \nonce_R,\nonce_T,\nonce_R' : \eq{\combine(\nonce_R,\nonce_T)}{\combine(\nonce_R',s(\nonce_R,\nonce_T,\nonce_R'))}\big)
\label{eq:lapkuntraceability}
\end{equation}
This condition is satisfied if $\combine$ is the xor operation (choose $s(\nonce_R,\nonce_T,\nonce_R') = \nonce_R' \oplus \nonce_T \oplus \nonce_R'$).
\[
\begin{array}{lll}
1:& E \longrightarrow T_A : \nonce_R\\
2:& T_A \longrightarrow E : \pair{\nonce_T}{\hash(\combine(\nonce_R,\nonce_T),\key_\textsf{A})} \\
&\null\hfill\\
3:& R \longrightarrow E : \nonce_R' \\
4:& E \longrightarrow R : \pair{s(\nonce_R,\nonce_T,\nonce_R')}{\hash(\combine(\nonce_R,\nonce_T),\key_\textsf{A})} \\
5:& R \longrightarrow E : \hash\big(\combine(\hash(\combine(\nonce_R,\nonce_T),\key_\textsf{A}),\nonce_R),\key_\textsf{A}\big)
\end{array}\]
The attacker starts by sending a name $\nonce_R$ to the tag, and gets the name $\nonce_T$ chosen by the tag as well as the hash $\hash(\combine(\nonce_R,\nonce_T),\key_\textsf{A})$. Then the attacker initiates a second round of the protocol with the reader. The reader sends first  a name $\nonce_R'$. The attacker is then able to answer, re-using the hash $\hash(\combine(\nonce_R,\nonce_T),\key_\textsf{A})$ sent by the tag in the first round,  choosing $s(\nonce_R,\nonce_T,\nonce_R')$ as a
replacement of the name $\nonce_T'$. Using Equation~(\ref{eq:lapkuntraceability}), there is a non negligible probability for the reader to accept the forged message as genuine.

This attack can be prevented by requiring $\combine$ to be injective on its first argument:
\begin{equation*}
  \forall a,b,x,y.\; \eq{\combine(a,b)}{\combine(x,y)} \Rightarrow \eq{a}{x}
\end{equation*}

\paragraph{Second Attack:} We have an unlinkability attack if we can distinguish between the answers of the tags, even though the hash function is assumed to be a PRF. This is possible if there exists a constant $g_1$ and a function $s$ such that:
\begin{equation}
 \Pr \big( x, y :  \combine(g_1,x) = \combine(s(x),y)\big)  \text{ is not negligible}
\label{eq:lapkuntraceabilityattack}
\end{equation}
If this is the case, then the unlinkability attack described 
below has a non negligible probability of success in distinguishing two consecutive rounds with the same tag $\textsf{A}$ from one round with the tag $\textsf{A}$ and one round with the tag ${\textsf{B}}$. 

The attack works as follows: it starts by impersonating the reader, sends $g_1$ to the tag and gets the response $\pair{\nonce_T}{\hash(\combine(g_1,\nonce_T),\key_\textsf{A})}$. Then the attacker initiates a new round of the protocol by sending $ s(\nonce_T)$ to the second tag. Using Equation~\ref{eq:lapkuntraceabilityattack},  there is a non negligible probability that the hash in the response from the tag $\textsf{A}$ in the second round of the protocol is the same as in the first round, whereas this will not be the case if the second round is initiated with ${\textsf{B}}$.
\begin{figure*}[htb]
  \noindent\framebox{\parbox{0.98\linewidth}{
      \[
        \begin{array}{lcl|lcl}
          1: E \longrightarrow T_{\textsf{A}} &:& g_1 & E \longrightarrow T_{\textsf{A}} &:& g_1\\
          2: T_{\textsf{A}} \longrightarrow E &:& \pair{\nonce_T}{\hash(\combine(g_1,\nonce_T),\key_\textsf{A})} & T_{\textsf{B}} \longrightarrow E &:& \pair{\nonce_T}{\hash(\combine(g_1,\nonce_T),\key_\textsf{A})}\\
                               &&\null\hfill\\
          3: E \longrightarrow T_{\textsf{A}} &:& s(\nonce_T) & E \longrightarrow T_{\textsf{B}} &:& s(\nonce_T)\\
          4: T_{\textsf{A}} \longrightarrow E &:& \pair{\nonce_T'}{\hash(\combine(s(\nonce_T),\nonce_T'),\key_\textsf{A})} & T_{\textsf{B}} \longrightarrow E &:& \pair{\nonce_T'}{\hash(\combine(s(\nonce_T),\nonce_T'),\key_\textsf{B})}
        \end{array}
      \]
    }}
  \caption{\label{fig:lakpuntraceability} Unlinkability attack against \lakp}
\end{figure*}

This attack can be prevented by asking $\combine$ to be right injective on its second argument:
\[ 
  \forall a,b,x,y.\; \eq{\combine(a,b)}{\combine(x,y)} \Rightarrow \eq{b}{y}
\]

\paragraph{Unlinkability of the \lakp protocol}
To prevent all the attacks against \lakp described above, we are going to require $\combine$ to be right and left injective. This can easily be expressed in the logic using the two axioms in Figure~\ref{fig:caxioms}, which are satisfied, for instance, when
$\combine$ is a pairing.
\begin{figure}[H]
\[\begin{array}{l}
\ite{\eq{u}{u'}}{\false}{\eq{\combine(u,v)}{\combine(u',v')}} = \false\\
\ite{\eq{v}{v'}}{\false}{\eq{\combine(u,v)}{\combine(u',v')}} = \false
\end{array}\]
\caption{\label{fig:caxioms}Injectivity axioms on the combination function $\combine$}
\end{figure}

Three messages are sent in a complete session of the \lakp protocol: two by the reader and one by the tag. Therefore, if we want to show interesting  properties of the \lakp protocol, we need to consider at least 6 terms in the trace (two full sessions, e.g. twice with the same tag $T_\textsf{A}$ or with the tag $T_\textsf{A}$ and the  tag $T_\textsf{B}$). This leads us to consider the $6$-Fixed Trace Privacy of the \lakp protocol.
\begin{theorem}
\label{thm:lakpuntrace} 
The \lakp protocol verifies $6$-Fixed Trace Privacy. In particular, the following formula is derivable using the axioms from Section~\ref{section:axioms} and the axioms in Figure~\ref{fig:caxioms}:
\[ \nonce_R,s^{\textsf{A}}_{\phi_0}, t^{\textsf{A}}_{\phi_1},\nonce_R',s'^{\textsf{A}}_{\phi_2}, t'^{\textsf{A}}_{\phi_3} \sim \nonce_R, s^{\textsf{A}}_{\phi_0},  t^{\textsf{A}}_{\phi_1},  \nonce_R', s'^{\textsf{B}}_{\phi_2},  t'^{\textsf{B}}_{\tilde\phi_3}\]
Where:
\begin{align*}
s^{\textsf{Id}}_\phi &= \pair{\nonce_T}{\hash(\combine(g(\phi),\nonce_T),\key_\textsf{\textsf{Id}})}\\
t^{\textsf{Id}}_\phi &=
                       \begin{array}[t]{l}
                         \itne{\beq{\hash\big(\combine(\nonce_R,\pi_1 (g(\phi))), \key_\textsf{\textsf{Id}}\big)}{\pi_2 (g(\phi))}\\}
                         {\hash\big(\combine(\pi_2( g(\phi)),\nonce_R),\key_\textsf{\textsf{Id}}\big)}
                       \end{array}\\
\phi_0& = \nonce_R \qquad\qquad
\phi_1 = \nonce_R,s^{\textsf{A}}_{\phi_0} \qquad\qquad
\phi_2 =\nonce_R,s^{\textsf{A}}_{\phi_0}, t^{\textsf{A}}_{\phi_1}\\
\phi_3& = \nonce_R,s^{\textsf{A}}_{\phi_0}, t^{\textsf{A}}_{\phi_1},\nonce_R',s'^{\textsf{A}}_{\phi_2} \qquad
\tilde \phi_3 = \nonce_R,s^{\textsf{A}}_{\phi_0}, t^{\textsf{A}}_{\phi_1},\nonce_R',s'^{\textsf{B}}_{\phi_2}
\end{align*}
\end{theorem}
The proof of this theorem is given in Appendix~\ref{appendix:lakp}. As for the \kclp protocol, by induction on $m$, it should be possible to generalize the result to an arbitrary $m$-Fixed Trace Privacy (if $m$ is independent of the security parameter), although we did not complete this generalization formally.



\section{Pseudo Random Number Generator}
\label{section:prng}
\label{section:pseudo-random}

A \cprng uses an internal state, which is updated at each call, and outputs a pseudo random number. This can be modeled by a function $G$ taking the internal state as input, and outputing a pair with the new internal state and the generated pseudo random number (retrieved using the projections $\pi_S$ and $\pi_o$). Besides, a function $\inits$ is used to initialized the internal state with a random seed (which can be hard-coded in the tag).
\begin{definition} 
\label{def:cprng}
A \cprng is a tuple of polynomial functions $(G,\inits)$ such that for every PPT adversary $\mathcal{A}$ and for 
every $n$, the following quantity is negligible in $\eta$:
\begin{multline*}
\big|\Pr\left(r\in \{0,1\}^\eta: \;\mathcal{A}(\pi_o(s_0),\dots,\pi_o(s_n)) = 1\right) -\\
 \Pr\left(r_0,\dots,r_n \in \{0,1\}^\eta:\;\mathcal{A}(r_0,\dots,r_n) = 1\right) \big|
\end{multline*}
where $s_0 = G(\inits(r,1^{\eta}))$ and for all $0 \le i < n,\; s_{i+1} = G(\pi_S(s_{i}))$.
\end{definition}

This can be translated in the logic by the $PRNG_n$ axioms:\\
\noindent\framebox{\parbox{0.98\linewidth}{
\[
\pi_o(s_0),\dots,\pi_o(s_n) \sim \nonce_0,\dots,\nonce_n 
\]
where $s_0 \equiv G(\inits(\nonce))$ and $\forall 0 \le i < n,\; s_{i+1} \equiv G(\pi_S(s_{i}))$.
}}
\vspace{0.3em}

The soundness of these axioms is an immediate consequence of Definition~\ref{def:cprng}:
\begin{proposition} 
The $(PRNG_n)_n$ axioms are sound in any computational model \cmodel where $(G,\inits)$ is interpreted as a \cprng.
\end{proposition}

For each protocol where a strict separation exists between the cryptographic material used for random number generation and the other primitives (e.g. encryption keys), pseudo random numbers generated using a \cprng can be abstracted as random numbers using the following proposition 
\ifdraft
(the proof can be found in Appendix~\ref{appendix:prng-proof}):
\else
(the proof can be found in~\cite{full-version}):
\fi
\begin{proposition} \label{prop:prng-sep}
For every names $\nonce,(\nonce_i)_{i \le n}$ and contexts $u_0,\dots,u_n$ that do not contain these names, the following formula is derivable using the axioms in Figure~\ref{figure:generic-axioms} and $PRNG_n$:
\[
u_0[\pi_o(s_0)], \dots, u_n[\pi_o(s_n)] \sim u_0[\nonce_0], \dots, u_n[\nonce_n] 
\]
where $s_0 \equiv G(\inits(\nonce))$ and $\forall 0 \le i < n,\; s_{i+1} \equiv G(\pi_S(s_{i}))$.
\end{proposition}

\begin{remark}[Forward Secrecy] We did not study forward secrecy of RFID protocols, but this could easily be done. The standard forward secrecy assumption on a \cprng states that leaking the internal state $\pi_S(s_{n})$ of the \cprng (e.g. with a physical attack on the RFID chip) does not allow the adversary to gain any information about the previously generated names $(\pi_o(s_{n}))_{i \le n}$. This could be expressed in the logic using, for example, the following formula:\\
\noindent\framebox{\parbox{0.98\linewidth}{
\[
\pi_o(s_0),\dots,\pi_o(s_n),\pi_S(s_{n}) \sim \nonce_0,\dots,\nonce_n,\pi_S(s_{n})
\]
where $s_0 \equiv G(\inits(\nonce))$ and $\forall 0 \le i < n,\; s_{i+1} \equiv G(\pi_S(s_{i}))$.
}}
\end{remark}


\section{Conclusion}
We gave a framework for formally proving the security of RFID protocols in the computational model: we expressed cryptographic assumptions on hash functions and an unlinkability property as formulas of the Complete Symbolic Attacker logic. We then illustrated this method on two examples, providing formal security proofs. We also showed that the security assumptions used in the proofs of these two protocols cannot be weakened (at least not in an obvious way).

What our framework is missing is an automatic tool for the logic, since the formal proofs are already heavy for simple protocols. Building such a tool would help streamline the design of formally verified protocols, and is the goal of our future research.

Compiling the process equivalence in our logic has already been explained in~\cite{bana14ccs}. In principle, we could use any automatic first order theorem prover to complete the proofs. However, the search space may be too large on our examples. This is why the focus of our current research is on the design of appropriate strategies.

\bibliographystyle{plain}
\bibliography{security}

\appendices

\ifdraft
\section{Proof of Proposition~\ref{proposition:generic:axioms}}
 \label{appendix:generic-axioms}
If terms $u$ and $v$ have no name in common then the distribution obtained when interpreting $u,v$ is the product of the distribution of $u$ and the distribution $v$. This generalized to interpretations under substitution $\sigma$ fixing all names common to $u$ and $v$.
\begin{proposition}
If $\names(u) \cap \names(v) \subseteq \dom(\sigma)$ then $\llbracket \vec u,\vec v \rrbracket_{\eta}^{\sigma} = \llbracket \vec u \rrbracket_{\eta}^{\sigma} \times \llbracket \vec v \rrbracket_{\eta}^{\sigma}$.
\end{proposition}

If there is a set of names $S$ such that the distributions obtained when interpreting $u$ and $v$ are the same for all valuation of the names in $S$ then distributions obtained when interpreting $u$ and $v$ are the same.
\begin{proposition}
If $\dom(\sigma) \cap S = \emptyset$ and if for all $\sigma'$ such that $dom(\sigma') = S$ we have $\llbracket \vec u \rrbracket_{\eta}^{\sigma \cup \sigma'} = \llbracket \vec v \rrbracket_{\eta}^{\sigma \cup \sigma'}$ then we have $\llbracket \vec u \rrbracket_{\eta}^{\sigma} = \llbracket \vec v \rrbracket_{\eta}^{\sigma}$.
\end{proposition}

We use the two proposition above to prove the soundness of the $Indep$ and $EqIndep$ axioms:
\begin{itemize}
\item $Indep$: Assume that for all computational model $\cmodel$ we have $\cmodel \models  \vec u \sim \vec v$. Using the $FreshNonce$ axiom we know that $\cmodel \models  \vec u,\nonce \sim \vec v,\nonce$. Therefore for all computational adversary $\mathcal{A}$ the following quantity is negligible:
\[
  \left| \Pr\left(\rho :\mathcal{A}(\llbracket \vec u, \nonce \rrbracket_{\eta,\rho}\right) = 1) - \Pr\left(\rho : \mathcal{A}(\llbracket \vec v, \nonce \rrbracket)_{\eta,\rho} = 1\right)\right| 
\] 

Let $\sigma$ be an arbitrary substitution fixing all names except $\nonce$. Then we have:
\begin{align*}
\llbracket \vec u, \nonce \oplus x \rrbracket_{\eta}^{\sigma} &= \llbracket \vec u\rrbracket_{\eta}^{\sigma} \times \llbracket \nonce \oplus x \rrbracket_{\eta}^{\sigma} \\
&= \llbracket \vec u\rrbracket_{\eta}^{\sigma} \times \left(\llbracket \nonce \rrbracket_{\eta}^{\sigma} \oplus \llbracket x \rrbracket_{\eta}^{\sigma} \right)\\
&= \llbracket \vec u\rrbracket_{\eta}^{\sigma} \times \llbracket \nonce \rrbracket_{\eta}^{\sigma}\\
&= \llbracket \vec u, \nonce \rrbracket_{\eta}^{\sigma}
\end{align*}

Since this holds for all $\sigma$ we have  $\llbracket \vec u, \nonce \oplus x \rrbracket_{\eta} = \llbracket \vec u, \nonce \rrbracket_{\eta}$. Similarly   $\llbracket \vec v, \nonce \oplus y \rrbracket_{\eta} = \llbracket \vec v, \nonce \rrbracket_{\eta}$. Therefore the following quantity is negligible:
\begin{multline*}
\big| \Pr\left(\rho :\mathcal{A}(\llbracket \vec u, x \xor\nonce \rrbracket_{\eta,\rho}\right) = 1) -\\
 \Pr\left(\rho : \mathcal{A}(\llbracket \vec v, y \xor\nonce \rrbracket)_{\eta,\rho} = 1\right)\big| 
\end{multline*}

\item $EqIndep$: For all computational model $\cmodel$:
\begin{align*}
&\;\Pr(\rho : \llbracket\eq{\nonce}{t}\rrbracket_{\eta,\rho} = \llbracket\false\rrbracket_{\eta,\rho})\\
= &\;\Pr(\rho : \llbracket\nonce\rrbracket_{\eta,\rho} = \llbracket t\rrbracket_{\eta,\rho})\\
= &\sum_{x \in \cdom_\sortmessage} \frac{1}{|\cdom_\sortmessage|}\Pr(\rho : \llbracket\nonce\rrbracket_{\eta,\rho} = x \wedge  \llbracket t\rrbracket_{\eta,\rho} = x)
\end{align*}
Since $\nonce$ does not appear in $t$ we have by independence:
\begin{align*}
=&\sum_{x \in \cdom_\sortmessage} \frac{1}{|\cdom_\sortmessage|}\Pr(\rho : \llbracket\nonce\rrbracket_{\eta,\rho} = x)  \times \Pr(\rho : \llbracket t\rrbracket_{\eta,\rho} = x)\\
= & \sum_{x \in \cdom_\sortmessage} \frac{1}{|\cdom_\sortmessage|} \times \frac{1}{2^\eta} \times \Pr(\rho : \llbracket t\rrbracket_{\eta,\rho} = x)\\
\le & \sum_{x \in \cdom_\sortmessage} \frac{1}{|\cdom_\sortmessage|} \times \frac{1}{2^\eta}\\
\le \;& \frac{1}{2^\eta}
\end{align*}
\end{itemize}


\section{Proof of Proposition~\ref{proposition:prfn}}
\label{appendix:prf}

We give here and show a simpler version of the Proposition~\ref{proposition:prfn}. The proof below can then easily be generalized.
\begin{proposition}\label{proposition-PRF}
The following axiom:\\
\noindent\framebox{\parbox{0.98\linewidth}{
\begin{multline*}
 \vec{u}, \ite{\eq{t_1}{t}}{\textbf{0}}{\hash(t,k)} \\\sim \vec{u}, \ite{\eq{t_1}{t}}{\textbf{0}}{\nonce} \quad (PRF_1)
\end{multline*}
where:
\begin{itemize}
\item the only occurrence of $\hash$ in $\vec{u},t$ is in a subterm $\hash(t_1,k)$
\item $\key$ does not occur in $t_1$ and only in key position in $t,\vec{u}$.
\item $\nonce$ is a name, that does not occur in $\vec{u},t,t_1$
\end{itemize}
}}
is computationally sound if $\hash(\cdot,k)$ is a PRF family.
\end{proposition}

\begin{IEEEproof}
Let $\cal A$ be an attacker that distinguishes the two sequences of terms. We construct an attacker on the PRF 
property as follows:
\begin{enumerate}
\item $\cal B$ draws all names appearing in $\vec{u}, t_1,t$, except $\key$. Let $\tau$ be the resulting sampling.
\item $\cal B$ computes (bottom-up) the distribution:
\[\sem{\vec{u},\ite{\eq{t_1}{t}}{\textbf{0}}{\hash(t,k)}}_{\tau_{+k,\rho}}\]
 calling the oracle each time there is an occurrence of $\hash(\cdot,k)$. ($\tau_{+k,\rho}$ is the extension of $\tau$ with a sampling of $\key$ and attacker's random tape $\rho$). This is possible, thanks to the occurrences assumptions on $\key$.
\item $\cal B$ simulates $\cal A$ on the result.
\end{enumerate}
Given a fixed $\tau$ we have:
\begin{align*}
&\prob\left(\key,\rho: \; {\cal B}^{{\cal O}_{\hash(\cdot,\key)}}(1^\eta)=1\right)=\\
&\prob\left(\key,\rho:\; {\cal A}(\sem{\vec{u},\ite{\eq{t_1}{t}}{\textbf{0}}{\hash(t,k)}}_{\tau_{+k,\rho}})=1\right)
\end{align*}
Hence,
\begin{multline}
\prob(\tau_{+k,\rho}: \; {\cal B}^{{\cal O}_{\hash(\cdot,\key)}}(1^\eta)=1) =\\  \prob(\tau_{+k,\rho}:\; {\cal A}(\sem{\vec{u},\ite{\eq{t_1}{t}}{\textbf{0}}{\hash(t,k)}}_{\tau_{+k,\rho}})=1) \label{eq:prf1}
\end{multline}

On the other hand, consider the term sequences, in which each occurrence $\hash(\cdot,\key)$ is replaced with a function symbol $g$: let $\vec{u'}, \ite{\eq{t_1}{t'}}{\textbf{0}}{g(t')}$ be the result of such a replacement in $\vec{u},\ite{\eq{t_1}{t}}{\textbf{0}}{\hash(t,k)} $. Let $\tau_{+g,\rho}$ be an extension of  $\tau$ with $g,\rho$,where $g$ is uniformly sampled from $\{0,1\}^* \rightarrow \{0,1\}^\eta$.
\begin{multline}
 C= \prob(\tau_{+k,\rho}: \; {\cal B}^{{\cal O}_{\hash(\cdot,\key)}}(1^\eta)=1) = \\
 \prob(\tau_{+g,\rho}:\; {\cal A}(\sem{\vec{u'},\ite{\eq{t_1}{t'}}{\textbf{0}}{g(t')}}_{\tau_{+g,\rho}})=1)
\label{eq:prf2}
\end{multline}
Hence, letting $\vec{U'}= \vec{u'},\ite{\eq{t_1}{t'}}{\textbf{0}}{g(t')}$,
\begin{multline*}
C=\\
\sum_{w\in \{0,1\}^\eta} \prob(\tau_{+g,\rho}: \; g(\sem{t_1}_{\tau_{+g,\rho}})=w) \hfill\\
\times \prob(\tau_{+g,\rho}: \; {\cal A}(\sem{\vec{U'}}_{\tau_{+g,\rho}})=1\quad  |\quad g(\sem{t_1}_{\tau_{+g,\rho}})=w)
\end{multline*}
Since $t_1$ is assumed not to contain any occurrence of $g$ and $g$ is drawn uniformly at random,
\begin{multline*}
C=\\\sum_{w\in \{0,1\}^\eta} \frac{1}{2^\eta}\times  \prob(\tau_{+g,\rho}: \; {\cal A}(\sem{\vec{U'}}_{\tau_{+g,\rho}})=1\quad  |\quad g(\sem{t_1}_\tau)=w)
\end{multline*}
Now, distinguishing cases,
\begin{multline*}
C= \sum_{w\in \{0,1\}^\eta} \frac{1}{2^\eta} ( \\
\prob(\tau_{+g,\rho}: \; \sem{t_1}_\tau = \sem{t'}_{\tau_{+g,\rho}} \wedge {\cal A}(\sem{\vec{u'}}_{\tau_{+g,\rho}},\textbf{0})=1 \\ 
\hfill  |\quad g(\sem{t_1}_\tau)=w)\\
+ \prob(\tau_{+g,\rho}: \; \sem{t_1}_\tau \neq \sem{t'}_{\tau_{+g,\rho}} \wedge  {\cal A}(\sem{\vec{u'},g(t')}_{\tau_{+g,\rho}})=1 \\
\hfill  |\quad g(\sem{t_1}_\tau)=w))
\end{multline*}
Now, since the only occurrences of $g$ in $t', \vec{u'}$ are in expressions $g(t_1)$, when we fix $g(\sem{t_1}_\tau)=w$, 
$\sem{\vec{u'}}_{\tau_{+g,\rho}}$ and $\sem{t'}_{\tau_{+g,\rho}}$ do not depend on $g,\rho$, but only on $w$: we write them
$\sem{\vec{u'}}_{\tau^w}$ and $\sem{t'}_{\tau^w}$ respectively. The above probability can then be written:

\begin{multline*}
C=\sum_{w\in \{0,1\}^\eta} \frac{1}{2^\eta} ( \\
\prob(\tau,\rho: \; \sem{t_1}_\tau = \sem{t'}_{\tau^w} \wedge   {\cal A}(\sem{\vec{u'}}_{\tau^w},\textbf{0})=1)\\
+ \prob(\tau_{+g,\rho}: \; \sem{t_1}_\tau \neq \sem{t'}_{\tau^w}\wedge  {\cal A}(\sem{\vec{u'}}_{\tau^w},\sem{g(t')}_{\tau_{+g,\rho}})=1  \\ 
\hfill  |\quad g(\sem{t_1}_\tau)=w))
\end{multline*}

Given $\tau, \rho,w$ and assuming $g(\sem{t_1}_\tau)=w$ and $\sem{t_1}_\tau\neq \sem{t'}_{\tau^w}$, the distribution
of $g(\sem{t'}_{\tau^w})$ is uniform; in other words, if ${\cal G}_{-w}$ is the set of functions $g$ from $\{0,1\}^*$ to $\{0,1\}^\eta$ such that $g(\sem{t_1}_\tau)=w$, for any $x\in \{0,1\}^*\setminus \{\sem{t_1}_\tau\}$, for any $y\in \{0,1\}^\eta$
\[ \prob(g \drawn{U}{{\cal G}_{-w}}:\; g(x)=y) =\frac{1}{2^\eta}\]
Therefore, given $\tau, w$ such that $\sem{t_1}_\tau=\sem{t'}_{\tau^w}$,
\begin{multline*}
\prob(\rho,g\drawn{U}{{\cal G}_{-w}}: \; {\cal A}(\sem{\vec{u'}}_{\tau^w},g(\sem{t'}_{\tau^w})=1))\\
= \prob(\rho,g\drawn{U}{{\cal G}_{-w}}, n\drawn{U}{\{0,1\}^\eta}: \; {\cal A}(\sem{\vec{u'}}_{\tau^w},n)=1))\\
\end{multline*}

Using this in the expression of $C$, we get:
\begin{multline*}  C= \sum_{w\in \{0,1\}^\eta} \frac{1}{2^\eta} ( \\
\prob(\tau,\rho: \; \sem{t_1}_\tau = \sem{t'}_{\tau^w} \wedge   {\cal A}(\sem{\vec{u'}}_{\tau^w},\textbf{0})=1)\\
+ \prob(\tau_{+g,\rho},n: \; \sem{t_1}_\tau \neq \sem{t'}_{\tau^w}\wedge  {\cal A}(\sem{\vec{u'}}_{\tau^w},n)=1  \\ 
\hfill  |\quad g(\sem{t_1}_\tau)=w))
\end{multline*}

If, extending $\tau$ with a sampling of $\key$,   $\sem{\hash(t_1,\key)}_{\tau_{+\key}}= w$, 
then $\sem{\vec{u'}}_{\tau^w}=\sem{\vec{u}}_{\tau_{+\key}}$ and $\sem{t'}_{\tau^w}=\sem{t}_{\tau_{+\key}}$.  
Since, now, $g$ only occurs in the conditional,
\begin{multline*}
C=\sum_{w\in \{0,1\}^\eta} \frac{1}{2^\eta} ( \\
\prob(\tau_{+\key},\rho: \; \sem{t_1}_\tau = \sem{t}_{\tau_{+\key}} \wedge   {\cal A}(\sem{\vec{u}}_{\tau_{+\key}},\textbf{0})=1\\
\hfill \quad |\quad \sem{\hash(t_1,\key)}_{\tau_{+\key}}=w)\\
+ \prob(\tau_{+\key},\rho,n: \; \sem{t_1}_\tau \neq \sem{t}_{\tau_{+\key}}\wedge  {\cal A}(\sem{\vec{u}}_{\tau_{+\key}},n)=1  \\ 
\hfill  |\quad \sem{\hash(t_1,\key)}_{\tau_{+\key}}=w))
\end{multline*}

Folding back, we get:
\begin{multline*}
C= \sum_{w\in \{0,1\}^\eta} \frac{1}{2^\eta} ( \\
\prob(\tau_{+\key},\rho,n: {\cal A}(\sem{\vec{u},\ite{\eq{t}{t_1}}{\textbf{0}}{n}}_{\tau_{+k,n}}) =1\\
\hfill  |\quad \sem{\hash(t_1,\key)}_{\tau_{+\key}}=w))
\end{multline*}

And summing :
\begin{align*}
& C= \\
&\prob(\tau_{+\key},\rho,n: {\cal A}(\sem{\vec{u},\ite{\eq{t}{t_1}}{\textbf{0}}{n}}_{\tau_{+k,n}}) =1)
\end{align*}

This, together with Equation~\ref{eq:prf1}, shows that the advantage of $\cal B$ is equal to the advantage of ${\cal A}$: we can break
the PRF property. A contradiction.
\end{IEEEproof}

\begin{remark}
In the proposition \ref{proposition-PRF}, the conditions on occurrences of $\key$ are necessary, unless we 
work in the random oracle model.
\end{remark}

The proofs of the $(PRF_n)$ axioms are identical to the previous one, fixing (instead of $w=g(t_1)$) the values $w_1,\ldots,w_n$ of $g(t_1),\ldots,g(t_n)$.


\else \fi

\section{Proof of Theorem~\ref{thm:lakpuntrace}}
\label{appendix:lakp}
%
 Unsurprisingly, it turns out that left and right injectivity of $\combine$ implies the injectivity of $\combine$: 
\begin{proposition} 
The following formula is derivable using the axioms from Section~\ref{section:axioms} and the axioms in Figure~\ref{fig:caxioms}:
\label{prop:inj}
\begin{align*}
\beq{\combine(u,v)}{\combine(u',v')} \;=\; &
  \ite{\eq{u}{u'}}
  {\\&\;\;\big(\ite{\eq{v}{v'}}{\true}{\false}\big)\\&}
  {\false} 
\end{align*}
\end{proposition}
The proof is straightforward rewriting using left and right injectivity and the $\ite{}{}{}$ axioms.

We are now ready to give the proof of Theorem~\ref{thm:lakpuntrace}. Most of the formulas are easy to prove, so we are going to focus on the formula explicitly given in the theorem statement, which is in our opinion the hardest case.

Before starting, we define several new terms in Figure~\ref{fig:lakp-terms}.
\begin{figure*}
\begin{align*}
&\alpha \equiv \pair{\nonce_T'}{\hash(\combine(g(\phi_2),\nonce'_T),\key_\textsf{A})}\\
&\beta \equiv \hash\big(\combine(\nonce'_R,\pi_1 (g(\phi_3))), \key_\textsf{A}\big)\\
&\gamma \equiv \hash\big(\combine(\pi_2( g(\phi_3)),\nonce'_R),\key_\textsf{A}\big)\\
&\epsilon_1  \equiv \beq{\nonce_R'}{g(\phi_0)} &
e_1 &\equiv \beq{\combine(\nonce'_R,\pi_1 (g(\phi_3)))}{\combine(g(\phi_0),\nonce_T)}&&\text{(in term $s_{\phi_0}^A$)}\\
&\epsilon_2  \equiv \beq{\nonce_R'}{\nonce_R}&
e_2 &\equiv \beq{\combine(\nonce'_R,\pi_1 (g(\phi_3)))}{\combine(\nonce_R,\pi_1 (g(\phi_1)))}&&\text{(in term $t_{\phi_1}^A$)}\\
&\epsilon_3  \equiv \beq{\nonce_R'}{\pi_2(g(\phi_1))}&
e_3 &\equiv \beq{\combine(\nonce'_R,\pi_1 (g(\phi_3)))}{\combine(\pi_2( g(\phi_1)),\nonce_R)}&&\text{(in term $t_{\phi_1}^A$)}\\
&\left.\begin{aligned}
\epsilon_4 & \equiv \beq{\nonce_R'}{g(\phi_2)}\\
\epsilon_4'&  \equiv \beq{\pi_1 (g(\phi_3))}{\nonce_T'}
\end{aligned}\right\}
&e_4 &\equiv \beq{\combine(\nonce'_R,\pi_1 (g(\phi_3)))}{\combine(g(\phi_2),\nonce'_T)}&&\text{(in term $s_{\phi_2}^A$)}\\
&\left.\begin{aligned}
\epsilon_5 & \equiv \beq{\nonce_R'}{\pi_1(g(\phi_3))}\\
\epsilon_5' & \equiv \beq{\nonce_R'}{\pi_2(g(\phi_3))}\\
\end{aligned}\right\}
&e_5 &\equiv \beq{\combine(\nonce'_R,\pi_1 (g(\phi_3)))}{\combine(\pi_2( g(\phi_3)),\nonce'_R)} &&\text{(in term $t_{\phi_3}^A$)}
\end{align*}
\caption{\label{fig:lakp-terms} Term Definitions for the \lakp Unlinkability Proof}
\end{figure*}

We have similar definition for the tilded versions $\tilde \alpha, \tilde \beta, \tilde \gamma,\dots$. We start by applying the $FA$ axiom several times:
\begin{prooftree}
\ax{\phi_2, \alpha, \beta, \gamma \sim  \phi_2, \tilde \alpha, \tilde \beta, \tilde \gamma}{}
\uinf{\phi_3, t'^A_{\phi_3} \sim \tilde \phi_3,  t'^B_{\tilde \phi_3}}{FA^*}
\end{prooftree}

 We are now going to use the $CS$ axiom on the conditional $e_4,e_5$ to split the proof. To do so we introduce the term:
\begin{multline*}
  u^x \equiv \ite{e_4}{\big(\ite{e_5}{x}{x}\big)\\}{\big(\ite{e_5}{x}{x}\big)}
\end{multline*}
and the terms:
\begin{align*}
u^x_1 &\equiv \ite{e_4}{0}{\big(\ite{e_5}{0}{x}\big)}\\
u^x_2 &\equiv \ite{e_4}{0}{\big(\ite{e_5}{x}{0}\big)}\\
u^x_3 &\equiv \ite{e_4}{\big(\ite{e_5}{0}{x}\big)}{0}\\
u^x_4 &\equiv \ite{e_4}{\big(\ite{e_5}{x}{0}\big)}{0}
\end{align*}

Similarly we introduced the tilded versions of these terms. We observe that for all term $s$ we have $s = u^s$ and $s = \tilde u^s$. Therefore we can apply the $CS$ axiom, which gives us:
\begin{prooftree}
\ax{\forall i \in \{ 1,\dots,4\},\;\;
\begin{array}[c]{l}
  \phi_2,e_4,e_5, u_i^\alpha, u_i^\beta, u_i^\gamma \\
  \qquad\sim  \phi_2,\tilde e_4,\tilde e_5, \tilde u_i^{\tilde \alpha}, \tilde u_i^{\tilde \beta}, \tilde u_i^{\tilde \gamma}
\end{array}
}
\uinf{\phi_2, \alpha, \beta, \gamma \sim  \phi_2, \tilde \alpha, \tilde \beta, \tilde \gamma}{CS^*}
\end{prooftree} 
We let $\phi = \phi_2,e_4,e_5$ and $\tilde \phi = \phi_2,\tilde e_4,\tilde e_5$.

\begin{figure*}
\textbf{Proof Tree $P_1$:}
\begin{prooftree}
\ax{\phi,  u_1^\alpha, u_1^\beta, u_1^\gamma \sim  \phi , u_1^\alpha, u_1^\nonce, u_1^\gamma}
\ax{\phi , u_1^\alpha, u_1^\gamma \sim\tilde \phi , \tilde u_1^{\tilde \alpha}, \tilde u_1^{\tilde \gamma}}
\uinf{\phi , u_1^\alpha, \nonce, u_1^\gamma \sim\tilde \phi , \tilde u_1^{\tilde \alpha}, \nonce, \tilde u_1^{\tilde \gamma}}{FreshNonce}
\uinf{\phi , u_1^\alpha, u_1^\nonce, u_1^\gamma \sim\tilde \phi , \tilde u_1^{\tilde \alpha}, \tilde u_1^\nonce, \tilde u_1^{\tilde \gamma}}{FA^*}
\ax{\tilde \phi , \tilde u_1^{\tilde \alpha}, \tilde u_1^\nonce, \tilde u_1^{\tilde \gamma} \sim  \tilde \phi , \tilde u_1^{\tilde \alpha}, \tilde u_1^{\tilde \beta}, \tilde u_1^{\tilde \gamma}}
\binf{\phi , u_1^\alpha, u_1^\nonce, u_1^\gamma \sim  \tilde \phi , \tilde u_1^{\tilde \alpha}, \tilde u_1^{\tilde \beta}, \tilde u_1^{\tilde \gamma}}{Trans}
\binf{\phi , u_1^\alpha, u_1^\beta, u_1^\gamma \sim  \tilde \phi , \tilde u_1^{\tilde \alpha}, \tilde u_1^{\tilde \beta}, \tilde u_1^{\tilde \gamma}}{Trans}
\end{prooftree} 
\textbf{Proof Tree $P_2$:}
\begin{prooftree}
\ax{\phi,  u_1^\alpha, u_1^\gamma  \sim  \phi , u_1^\alpha, u_1^\nonce}
\ax{\phi , u_1^\alpha \sim\tilde \phi , \tilde u_1^{\tilde \alpha}}
\uinf{\phi , u_1^\alpha, \nonce  \sim\tilde \phi , \tilde u_1^{\tilde \alpha}, \nonce}{FreshNonce}
\uinf{\phi , u_1^\alpha, u_1^\nonce  \sim\tilde \phi , \tilde u_1^{\tilde \alpha}, \tilde u_1^\nonce}{FA^*}
\ax{\tilde \phi , \tilde u_1^{\tilde \alpha}, \tilde u_1^\nonce  \sim  \tilde \phi , \tilde u_1^{\tilde \alpha}, \tilde u_1^{\tilde \gamma}}
\binf{\phi , u_1^\alpha, u_1^\nonce  \sim  \tilde \phi , \tilde u_1^{\tilde \alpha}, \tilde u_1^{\tilde \gamma} }{Trans}
\binf{\phi , u_1^\alpha, u_1^\gamma  \sim  \tilde \phi , \tilde u_1^{\tilde \alpha}, \tilde u_1^{\tilde \gamma}}{Trans}
\end{prooftree} 
\caption{\label{fig:lakp-prooftree} Derivations $P_1$ and $P_2$}
\end{figure*}

\begin{itemize}
\item \textbf{Case $i=1$:}  Let $\nonce$ be a fresh name, we start by the derivation $P_1$ displayed in Figure~\ref{fig:lakp-prooftree}.
Using $EqIndep$ we know that $\epsilon_1 = \epsilon_2 = \epsilon_3  = \false$, and using the left injectivity of $\combine$ this shows that $e_1 = e_2 = e_3 = \false$. Therefore we know that:
\begin{multline*}
  u_1^\beta = v^\beta \equiv \ite{e_1}{0}{\ite{e_2\\
    }{0}{\big( \ite{e_3}{0}{\big(u_1^\beta\big)}\big)}}
\end{multline*}
\begin{multline*}
  u_1^\nonce = {v^\nonce} \equiv\ite{e_1}{0}{\ite{e_2\\
    }{0}{\big( \ite{e_3}{0}{\big(u_1^\nonce\big)}\big)}}
\end{multline*}
Hence we can apply the $PRF$ axiom, which shows that:
\begin{prooftree}
\ax{}
\uinf{\phi,  u_1^\alpha, v^\beta, u_1^\gamma \sim  \phi , u_1^\alpha, v^\nonce, u_1^\gamma}{PRF}
\uinf{\phi,  u_1^\alpha, u_1^\beta, u_1^\gamma \sim  \phi , u_1^\alpha, u_1^\nonce, u_1^\gamma}{Congr}
\end{prooftree}

Similarly we show that:
\begin{prooftree}
\ax{}
\uinf{ \tilde\phi , \tilde u_1^{\tilde\alpha}, \tilde v^\nonce, \tilde u_1^{\tilde\gamma} \sim \tilde \phi,  \tilde u_1^{\tilde\alpha}, \tilde v^{\tilde\beta}, \tilde u_1^{\tilde\gamma} }{PRF}
\uinf{  \tilde\phi , \tilde u_1^{\tilde\alpha}, \tilde u_1^\nonce, \tilde u_1^{\tilde\gamma} \sim \tilde \phi,  \tilde u_1^{\tilde\alpha}, \tilde u_1^{\tilde\beta}, \tilde u_1^{\tilde\gamma}}{Congr}
\end{prooftree}

It remains to show that $\phi , u_1^\alpha, u_1^\gamma \sim\tilde \phi , \tilde u_1^{\tilde \alpha}, \tilde u_1^{\tilde \gamma}$. We do this exactly like we did to get rid of the $u_1^\beta$ and $\tilde u_1^{\tilde \beta}$. First we use $FA$, $Trans$ and $FreshNonce$ to get the derivation $P_2$ displayed in Figure~\ref{fig:lakp-prooftree}.

The formulas $\phi,  u_1^\alpha, u_1^\gamma  \sim  \phi , u_1^\alpha, u_1^\nonce$ and $\tilde \phi , \tilde u_1^{\tilde \alpha}, \tilde u_1^\nonce  \sim  \tilde \phi , \tilde u_1^{\tilde \alpha}, \tilde u_1^{\tilde \gamma}$ are dealt with exactly like we did for $\phi,  u_1^\alpha, u_1^\beta, u_1^\gamma \sim  \phi , u_1^\alpha, u_1^\nonce, u_1^\gamma$, introducing the corresponding conditional tests. We do not detail these two cases, but notice that the right injectivity of $\combine$ is needed for them.

We now need to show that $\phi , u_1^\alpha \sim\tilde \phi , \tilde u_1^{\tilde \alpha}$, which is done by applying the $FA$ axiom several time:
\begin{prooftree}
\ax{
\begin{array}{l}
\phi_2,\nonce_R',\nonce_T',\hash(\combine(g(\phi_2),\nonce'_T),\key_\textsf{A})\\
\qquad\qquad\sim \phi_2,\nonce_R',\nonce_T',\hash(\combine(g(\phi_2),\nonce'_T),\key_\textsf{B})
\end{array}
}{}
\uinf{\phi_3 \sim \tilde \phi_3}{FA^*}
\uinf{\phi , u_1^\alpha \sim\tilde \phi , \tilde u_1^{\tilde \alpha}}{FA^*}
\end{prooftree}
Let $\psi \equiv \phi_2,\nonce_R',\nonce_T'$, it is then easy to show that $\psi,\hash(\combine(g(\phi_2),\nonce'_T),\key_\textsf{A}) \sim \psi,\hash(\combine(g(\phi_2),\nonce'_T),\key_\textsf{A})$ is derivable using the fact that $\nonce'_T$ is fresh in $\psi$, the right injectivity of $\combine$ and the $PRF$ axiom.

\item \textbf{Case $i=2$ and $i=3$:} These case are very similar to the case $i=1$, except that we need to use the $Dup$ axiom at some point to get rid of the double occurrence of $\gamma$ (in case $i=2$) or $\alpha$ (in case $i=3$).

\item \textbf{Case $i=4$:} Using Proposition~\ref{prop:inj} we know that
\begin{gather*}
e_4 = \ite{\epsilon_4}{\big(\ite{\epsilon_4'}{\true}{\false}\big)}{\false}\\
e_5 = \ite{\epsilon_5}{\big(\ite{\epsilon_5'}{\true}{\false}\big)}{\false}
\end{gather*}

Since booleans $\epsilon_4'  \equiv \beq{\pi_1 (g(\phi_3))}{\nonce_T'}$ and $\epsilon_5  \equiv \beq{\nonce_R'}{\pi_1(g(\phi_3))}$ we have:
\begin{align*}
  &\ite{\epsilon_4'}{\big( \ite{\epsilon_5}{\true}{\false}\big)}{\false}\\
  &= \ite{\epsilon_4'}{
    \left(
    \ite{\beq{\nonce_R'}{\nonce_T'}
    \begin{array}[t]{l}
      }{\true\\
      }{\false}
    \end{array}
    \right)
    }{\false}\\
  &= \ite{\epsilon_4'}{\big( \ite{\false}{\true}{\false}\big)}{\false}\\
  &= \false
\end{align*}
And therefore, for all term $v$ we have $u^v_4 = 0$. Similarly we have $\tilde u^v_4 = 0$ This means that we have:
\begin{prooftree}
\ax{\phi_3 \sim \tilde \phi_3}
\uinf{\phi , 0,0,0 \sim  \tilde \phi ,0,0,0}{FA}
\uinf{\phi , u_4^\alpha, u_4^\beta, u_4^\gamma \sim  \tilde \phi , u_4^{\tilde \alpha}, u_4^{\tilde \beta}, u_4^{\tilde \gamma}}{Congr}
\end{prooftree} 
We already showed in the case $i=1$ that $\phi_3 \sim \tilde \phi_3$ is derivable.
\end{itemize}


\ifdraft
\section{Proof of Proposition~\ref{prop:prng-sep}}
\label{appendix:prng-proof}
Let $\nonce,(\nonce_i)_{i \le n}$ and $u_0,\dots,u_n$ be such that $u_0,\dots,u_n$ do not contain $\nonce,(\nonce_i)_{i \le n}$. Let $s_0 \equiv G(\inits(\nonce))$ and $\forall 0 \le i < n,\; s_{i+1} \equiv G(\pi_S(s_{i}))$. We want to give a derivation of:
\[
u_0[\pi_o(s_0)], \dots, u_n[\pi_o(s_n)] \sim u_0[\nonce_0], \dots, u_n[\nonce_n] 
\]
using the axioms in Figure~\ref{figure:generic-axioms} and $PRNG_n$.

For all $i$, we let the context $C_i$ and the names $(\nonce_{i,j}^{p})_j$ be such that $u_i \equiv C_i[(\nonce_{i,j}^{p})_j]$ and $C_i$ does not contain any name (only function applications and holes). Then using the $FA$ axiom we have:
\begin{prooftree}
\ax{\left((\nonce_{i,j}^{p})_j\right)_{i \le n},\left(\pi_o(s_i)\right)_{i \le n} \sim \left((\nonce_{i,j}^{p})_j\right)_{i \le n},\left(\nonce_i\right)_{i \le n}}
\uinf{\left((\nonce_{i,j}^{p})_j,\pi_o(s_i)\right)_{i \le n} \sim \left((\nonce_{i,j}^{p})_j,\nonce_i\right)_{i \le n}}{Perm}
\uinf{\left(C_i[(\nonce_{i,j}^{p})_j][\pi_o(s_i)]\right)_{i \le n} \sim \left(C_i[(\nonce_{i,j}^{p})_j][\nonce_i]\right)_{i \le n}}{FA^*}
\end{prooftree}
Now, we can use the $Duplicate$ axiom to get rid of multiple occurrences of the same name: indeed if there exists a name $\textsf{m}$ such that $\textsf{m} \equiv \nonce_{i,j}^p$ and $\textsf{m} \equiv \nonce_{i',j'}^p$ then we can keep only one occurrence of $\textsf{m}$. Let $\textsf{m}_1,\dots,\textsf{m}_l$ be such that for all $i\ne j, \textsf{m}_i \ne \textsf{m}_j$ and $\{\textsf{m}_i \mid i\le l\} = \{\nonce^p_{i,j}\mid i \le n,j\}$, then:
\begin{prooftree}
\ax{\left(\textsf{m}_i\right)_{i \le n},\left(\pi_o(s_i)\right)_{i \le n} \sim \left(\textsf{m}_i\right)_{i \le n},\left(\nonce_i\right)_{i \le n}}
\uinf{\left((\nonce_{i,j}^{p})_j\right)_{i \le n},\left(\pi_o(s_i)\right)_{i \le n} \sim \left((\nonce_{i,j}^{p})_j\right)_{i \le n},\left(\nonce_i\right)_{i \le n}}{Dup^*}
\end{prooftree}
Now by assumptions we know that  $\{\nonce,(\nonce_i)_{i \le n}\} \cap \{\textsf{m}_i \mid i\le l\} = \emptyset$, therefore we can apply the $FreshNonce$ axiom for all $i \le l$ to get rid of $\textsf{m}_i$. Finally we conclude with the $PRNG_n$ axiom:
\begin{prooftree}
\ax{}
\uinf{\left(\pi_o(s_i)\right)_{i \le n} \sim \left(\nonce_i\right)_{i \le n}}{PRNG_n}
\uinf{\left(\textsf{m}_i\right)_{i \le n},\left(\pi_o(s_i)\right)_{i \le n} \sim \left(\textsf{m}_i\right)_{i \le n},\left(\nonce_i\right)_{i \le n}}{FreshNonce^*}
\end{prooftree}
\else \fi

\end{document}